\documentclass{article}
\pdfoutput=1

\usepackage{arxiv}

\usepackage[utf8]{inputenc}
\usepackage[T1]{fontenc}
\usepackage{hyperref}
\usepackage{url}
\usepackage{booktabs}
\usepackage{amsmath}
\usepackage{amsfonts}
\usepackage{microtype}
\usepackage{graphicx}
\usepackage{tabularx}
\usepackage{xltabular}
\usepackage{ragged2e}
\usepackage{caption}

\title{A unified power-grid representation for reuse across network structures and computational tasks}

\date{}

\usepackage{authblk}

\author[1,2]{Yanhao Huang}
\author[1,2]{Changqing Liu}
\author[1,2]{Zichang Wang}
\author[1,2]{Qianhong Wu}
\affil[1]{State Key Laboratory of Power Grid Safety, Beijing 100192, China}
\affil[2]{China Electric Power Research Institute, Beijing 100192, China}

\renewcommand{\shorttitle}{A unified power-grid representation}

\hypersetup{
pdftitle={A unified power-grid representation for reuse across network structures and computational tasks},
pdfauthor={Yanhao Huang, Changqing Liu, Zichang Wang, Qianhong Wu},
}

\begin{document}
\maketitle

\begin{abstract}
Data-driven power-system models are typically developed for specific grids and computational tasks, but their performance can deteriorate markedly or even fail when network structures or analytical objectives change. This paper develops a unified grid representation that separates physical-grid description from downstream computation. A self-supervised encoder represents each grid as a variable number of fixed-dimensional node, branch and global vectors under a common latent description. Although pretrained only on systems with at most 270 buses, the frozen encoder transfers without adaptation to a 70,000-bus network, more than 250 times larger in bus count, while preserving 0.7919 bus-correspondence accuracy. The same representation supports four independently trained downstream tasks: power-flow calculation, reactive-power adjustment, operating-condition generation and transient-stability assessment. On the previously unseen 70,000-bus grid, power-flow calculation achieves mean errors of 0.108$^{\circ}$ in phase angle and 3.8$\times$10\textsuperscript{-6} p.u. in voltage magnitude. Reactive-power adjustment achieved a 75\% PSASP-verified correction rate on the previously unseen 70,000-bus grid, while operating-condition generation delivered at least one PSASP-verified feasible state for 82.2\% of requests on previously unseen 400--1,000-bus grid families. For transient-stability assessment, the frozen encoder achieves accuracy within 0.8 percentage points of full encoder adaptation under matched training conditions. These results demonstrate that a unified power-grid representation can be learned once and reused unchanged across structurally different power systems and heterogeneous computational tasks.
\end{abstract}

\section*{Introduction}

Data-driven methods for power-system analysis and operation are increasingly expected to remain effective across a wide range of operating conditions, network topologies, contingencies and analysis tasks. The same physical grid may be studied under different generation--load patterns and network configurations and for substantially different purposes, including steady-state analysis, corrective control and dynamic-security assessment. Although these applications require different mathematical formulations and solution procedures, they repeatedly depend on common information about network connectivity, equipment characteristics and operating state. Artificial-intelligence methods have been applied to a broad range of power-system analysis and operation tasks\textsuperscript{[1]}, while recent power-grid foundation-model studies have begun to explore more reusable learning frameworks for such applications\textsuperscript{[2,3]}. Benchmark and large-scale simulation datasets similarly span multiple network configurations and operating conditions\textsuperscript{[4,5]}. Data-driven models, however, are still commonly developed and validated for specific systems, tasks and operating regimes. This creates a mismatch between the range of conditions encountered in practical power-system analysis and the comparatively narrow scope over which individual data-driven models are usually developed.

One response to this mismatch has been to improve the structural generalization of data-driven models across different power grids. Early power-grid foundation-model studies proposed combining self-supervised power-flow pretraining with graph-based models that can be adapted to different network topologies\textsuperscript{[6]}. More recent work has moved from this formulation towards explicit evaluation of structural transfer. Topology-transferable AC optimal-power-flow models have been trained across multiple network structures and evaluated on previously unseen grids and larger systems, with both zero-shot inference and target-grid adaptation considered\textsuperscript{[7]}. Physics-informed self-supervised graph pretraining has also been shown to improve generalization under fixed and changing network topologies\textsuperscript{[8]}. Other graph-based approaches have extended structural transfer across unseen topologies, different system scales and network reconfigurations\textsuperscript{[9-12]}, while studies of power flow, optimal power flow, contingency analysis and topology control have examined topology perturbations, N-1 and higher-order outages, cross-scale operation and out-of-distribution network states\textsuperscript{[13-18]}. Such structural changes are not merely changes in graph connectivity: they can materially affect voltage stability, power-transfer capability and system synchronization\textsuperscript{[19-21]}. Existing studies therefore establish substantial progress in structural transfer, but largely within particular task formulations or through protocols that may still include target-grid adaptation.

Beyond structural transfer, another line of research has explored whether learned models or representations can be reused across different power-system analysis tasks. In steady-state applications, generic representations learned from large-scale electricity time-series data have also been adapted to a broad set of forecasting, imputation, anomaly-detection and classification tasks\textsuperscript{[22]}, while reusable learned components have been developed for closely related power-flow-constrained computations\textsuperscript{[23]}. Other data-driven studies have addressed task-specific problems such as feasibility restoration and assessment and the generation of operating states, grid structures and renewable scenarios\textsuperscript{[24-30]}. In transient analysis, generalization has mainly been studied within dynamic-security tasks, where models are transferred across operating conditions, faults, contingencies, dynamic regimes and, in some cases, different power systems\textsuperscript{[31-36]}. Reviews and security-oriented data-generation studies nevertheless continue to identify sample coverage, model updating, reliability and adequate representation of operating points near the security boundary as important issues\textsuperscript{[37-40]}. Overall, existing studies demonstrate substantial reuse among related steady-state or common-modality tasks and substantial generalization within transient-analysis tasks, but provide much less evidence that the same grid representation can remain unchanged when reused across both steady-state and transient computations.

More directly related to the present question, several recent studies have combined structural generalization with reuse across multiple power-system tasks. A unified neural-solver framework further supports several steady-state computations, including power flow, optimal power flow and state estimation, while transfer to previously unseen grids relies on grid-specific fine-tuning\textsuperscript{[3]}. Physics-informed self-supervised graph pretraining has been used as a common initialization for several power-system analysis tasks under both fixed and changing network topologies, but the pretrained model is subsequently fine-tuned for each downstream task\textsuperscript{[8]}. A multiplex graph-Transformer framework shares a node encoder between static state estimation and AC power flow and evaluates the model on previously unseen topologies, although the shared encoder is subsequently unfrozen during multi-task fine-tuning\textsuperscript{[9]}. Heterogeneous graph foundation models have likewise been pretrained across multiple grid structures and system scales before being adapted to downstream optimal-power-flow-related tasks\textsuperscript{[10]}. Together, these studies show that structural variation and multi-task reuse can be addressed within a common learning framework. However, the shared representation is generally still shaped by downstream task- or grid-specific optimization, leaving open whether the same representation can remain unchanged across both dimensions.

Across these studies, a key methodological distinction is which part of the learned model is allowed to change after pretraining. Early power-grid foundation-model work envisaged self-supervised pretraining followed by task- or subgrid-specific fine-tuning\textsuperscript{[6]}. Subsequent studies have adopted different adaptation strategies: topology-transferable AC optimal-power-flow models have combined zero-shot evaluation with target-grid adaptation\textsuperscript{[7,41]}, physics-informed self-supervised graph pretraining has been followed by task-specific fine-tuning\textsuperscript{[8]}, shared encoders have been further optimized during multi-task training\textsuperscript{[9]}, and heterogeneous graph foundation models have considered full, partial and head-only downstream adaptation\textsuperscript{[10]}. Unified steady-state neural solvers likewise use grid-specific fine-tuning when transferred to unseen systems\textsuperscript{[3]}, while generic electricity time-series representations are adapted to downstream tasks\textsuperscript{[22]}. Across these representative frameworks, reuse is therefore generally accompanied by some form of parameter adaptation, even when a common pretrained model is available. This is consistent with broader foundation-model and graph representation-learning paradigms, in which general features are learned through large-scale pretraining, masked reconstruction, contrastive learning or local--global graph architectures and subsequently reused for downstream applications\textsuperscript{[42-45]}. Related work on compositional learning and domain generalization considers systematic reuse and generalization to unseen distributions, but does not require the underlying representation to remain unchanged\textsuperscript{[46,47]}. The unresolved question is therefore whether the grid representation itself can remain fixed across both unseen grids and heterogeneous downstream tasks, with task-specific specialization confined to external reasoners.

This distinction is particularly relevant to the practical deployment of data-driven methods in power-system analysis and operation. First, within the same power system, extreme events such as cascading outages or unanticipated contingencies can cause substantial and rapid changes in network topology. When the resulting network structure and operating state move far beyond those represented in the training data, a data-driven model developed for normal operating conditions or prescribed contingency sets may experience severe performance degradation and, in extreme cases, may no longer provide reliable results. Such topology changes can also materially alter voltage stability, power-transfer capability and synchronization characteristics, further increasing the difficulty of extrapolation from previously observed conditions. Second, when a trained model is applied to a power system that was not represented during model development, grid-specific retraining or fine-tuning is commonly required. Differences in network topology, system size, component parameters and operating characteristics can make direct transfer difficult and can lead to substantial performance degradation, while adaptation to each new grid requires additional data, model updating and validation. These requirements increase the cost of transferring data-driven models between power systems and can substantially limit their practical deployment. A representation that is tightly coupled to its development grid or downstream objective can therefore limit reuse even when transfer learning remains possible.

This gap raises a central question: \textbf{can a }\textbf{unified representation}\textbf{ of a power grid be learned once and reused unchanged across structurally different }\textbf{even previously unseen }\textbf{grids and heterogeneous downstream tasks}\textbf{? }Here, a unified representation refers to a common representation framework that can encode different grids and operating states without being redesigned for each system or task. Reused unchanged means that the representation is learned once, while subsequent grid- and task-specific computation is carried out without modifying the shared representation. The question is therefore whether a common grid representation can support both structural transfer and heterogeneous analyses spanning steady-state computation and transient-security assessment. Its role is to provide a common physical representation of the grid, while task-specific solution procedures remain specialized to their respective analytical objectives.

This paper develops a standardized representation framework for power grids of different sizes and topologies, together with a self-supervised encoder for unified grid representation that decouples physical-grid representation from downstream computation. Building on the direction outlined in [3] towards adapting grid foundation models to new tasks and grid topologies without fine-tuning, we realize this objective at the representation level: the grid is encoded through one fixed representation mechanism, while task-specific computation is learned separately by downstream reasoners. We examine whether this common representation preserves physically meaningful information, generalizes across changes in grid structure, and can be reused unchanged across four heterogeneous power-system tasks spanning steady-state computation and transient-security analysis: power-flow calculation, reactive-power adjustment, operating-condition generation and transient-stability assessment (TSA). Representation-level analyses, task-specific evaluations and independent physical verification are then used to characterize the robustness and limits of this representation reuse.

\section*{Methods}

\subsection*{2.1 Framework for unified grid representation and task-specific computation}

As shown in Fig. 1, the proposed framework separates a unified representation of the physical power grid from task-specific computation, allowing the same frozen encoder to be reused across different network structures and downstream tasks. The interface of Fig. 1 is read as follows. Each power-grid case is expressed as a variable-size graph and mapped by the encoder to addressable node, branch and global tokens in a common 192-dimensional latent space. The encoder contains eight GPS-style layers and 16 global tokens, and its weights, normalization statistics and token semantics are fixed after self-supervised pretraining. Task-specific reasoners then read the frozen representation together with information introduced only after encoding, including specified power-flow quantities, voltage-violation descriptors, requested operating conditions and fault information. The resulting outputs are mapped back to explicit power-system quantities and evaluated independently using Power System Analysis Software Package (PSASP).

\begin{figure}[htbp]
\centering
\includegraphics[width=\linewidth]{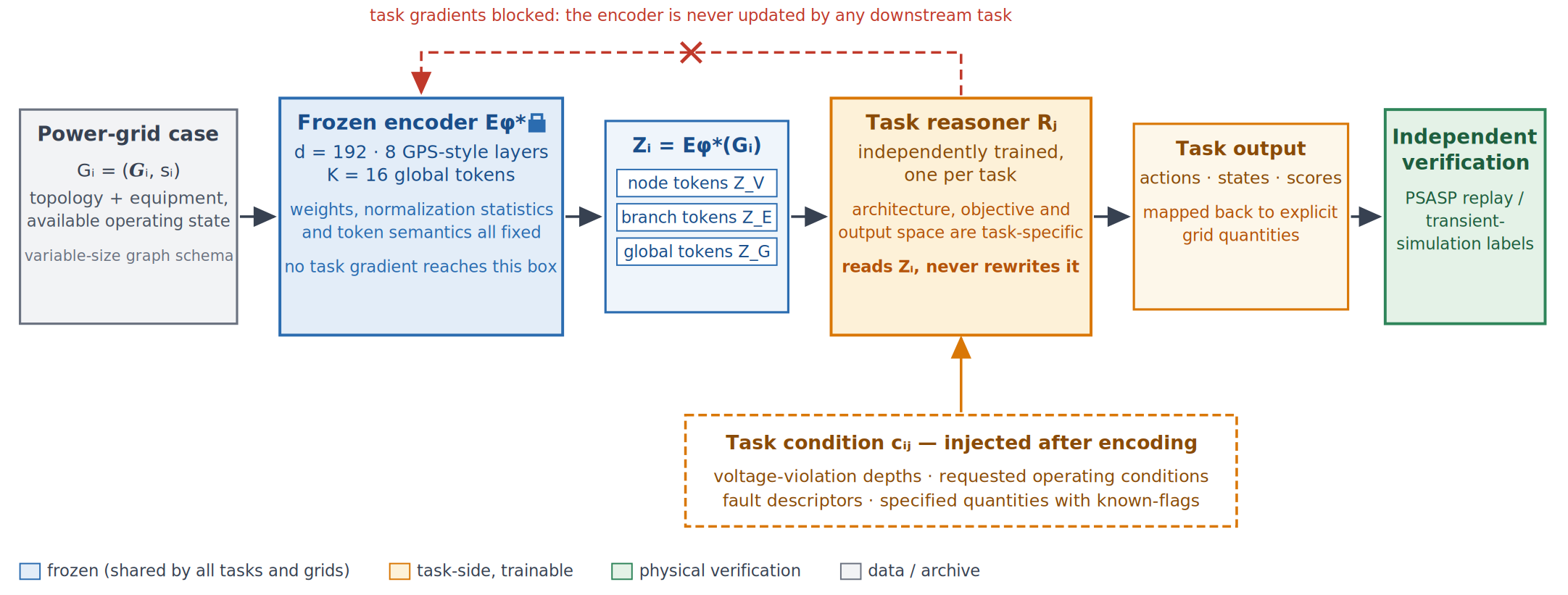}
\caption*{Fig. 1 | Framework for unified grid representation and task-specific computation.}
\end{figure}

For the grid-representation pathway, each power-system case is described by \textit{G}\textit{\textsubscript{i}}=($\mathcal{G}$\textit{\textsubscript{i}}, \textit{s}\textit{\textsubscript{i}}), where $\mathcal{G}$\textit{\textsubscript{i}} contains the network topology and equipment configuration and \textit{s}\textit{\textsubscript{i}} contains the available operating state. Grids of different sizes and topologies are expressed through the same variable-size graph schema and processed by the shared encoder \textit{E}\textit{\textsubscript{$\phi$}}\textit{\textsubscript{,}} producing an addressable latent representation \textit{Z}\textit{\textsubscript{i}} = \textit{E}\textit{\textsubscript{$\phi$}}(\textit{G}\textit{\textsubscript{i}}) with node-, branch- and system-level components. The number of represented elements varies with the grid, whereas the representation mechanism and latent feature dimension remain shared.

Downstream computation is separated from this representation pathway. For each computational task \textit{T}\textit{\textsubscript{j}}, an independently parameterized reasoner $R_{\psi_j}$ reads the frozen grid representation \textit{Z}\textit{\textsubscript{i}} together with an external task condition \textit{c}\textit{\textsubscript{i}}\textit{\textsubscript{j}} and produces the corresponding task output $\hat{y}_{ij} = R_{\psi_j}(Z_i, c_{ij})$. The task condition contains information required only by the corresponding computational objective, such as voltage-violation descriptors, requested operating conditions or fault information. Different tasks may therefore have different conditioning variables, reasoner architectures, outputs and training objectives while using the same grid representation. The resulting output is subsequently assessed by the corresponding numerical or simulator-based physical verifier.

The shared encoder is learned before downstream task training and is subsequently kept fixed. It is first optimized using self-supervised objectives without downstream task labels or task identities,

\begin{equation}
\phi^{*} = \arg\min_{\phi} L_{\mathrm{SSL}}(\phi)
\end{equation}

after which its parameters and associated representation interface are frozen. Each downstream reasoner is then trained independently while the encoder remains outside the task-specific optimizer,

\begin{equation}
\psi_{j}^{*} = \arg\min_{\psi_{j}} L_{j}(\psi_{j};\,\phi^{*})
\end{equation}

This sequential training procedure ensures that downstream task supervision neither determines nor modifies the shared grid representation. Across the grid dimension, the same encoder operates on networks with different sizes and topologies. Across the task dimension, the same frozen representation serves independently trained reasoners with different inputs, outputs, objectives and physical endpoints. We instantiate this framework for four power-system computations: power-flow calculation, reactive-power adjustment, operating-condition generation and TSA.

The representation architecture and self-supervised training are described in Section 2.2, the four task-specific reasoners and their physical verification procedures in Section 2.3, and the power-system data and simulations in Section 2.4.

\subsection*{2.2 Unified grid representation}

As illustrated in Fig. 2, power grids of different sizes and topologies are represented as variable-size graphs and mapped by a shared encoder into aligned node, directed-branch and global representations. Each physical bus is represented by a node record, while each physical branch is represented by two directed branch records. Structural information is augmented by Laplacian positional encodings (LaPE) and random-walk structural encodings (RWSE), and sample-local random identity channels are appended to node and branch records as within-case addressing cues. A separate directory maintains the correspondence between simulator objects and graph records, allowing latent representations and downstream outputs to remain associated with the corresponding buses, generators, shunts and branches. Node and branch inputs are projected into a common latent space with dimension  \textit{d} = 192, together with \textit{K}=16 learnable global tokens. The encoder contains eight GPS-style layers with four attention heads. Within each layer, local edge-aware message passing propagates information along network connections, after which the global tokens aggregate information from the node representations and return system-level context to the nodes through global attention. Residual feed-forward updates complete each layer. For a power-system case \textit{G}, the encoder produced \textit{Z} = \textit{E}\textit{\textsubscript{$\phi$}}(\textit{G}) = \{\textit{Z}\textit{\textsubscript{V}}, \textit{Z}\textit{\textsubscript{E}} , \textit{Z}\textit{\textsubscript{G}}\}, where \textit{Z}\textit{\textsubscript{V}}, \textit{Z}\textit{\textsubscript{E}} and \textit{Z}\textit{\textsubscript{G}} denote the node-, branch- and global-token representations, respectively. The numbers of node and branch tokens therefore vary with network size and topology, whereas the encoder parameters, latent dimension and token semantics remain common across grids. This structure preserves element-level addressability while allowing information to propagate through both local electrical connections and system-level context.

\begin{figure}[htbp]
\centering
\includegraphics[width=\linewidth]{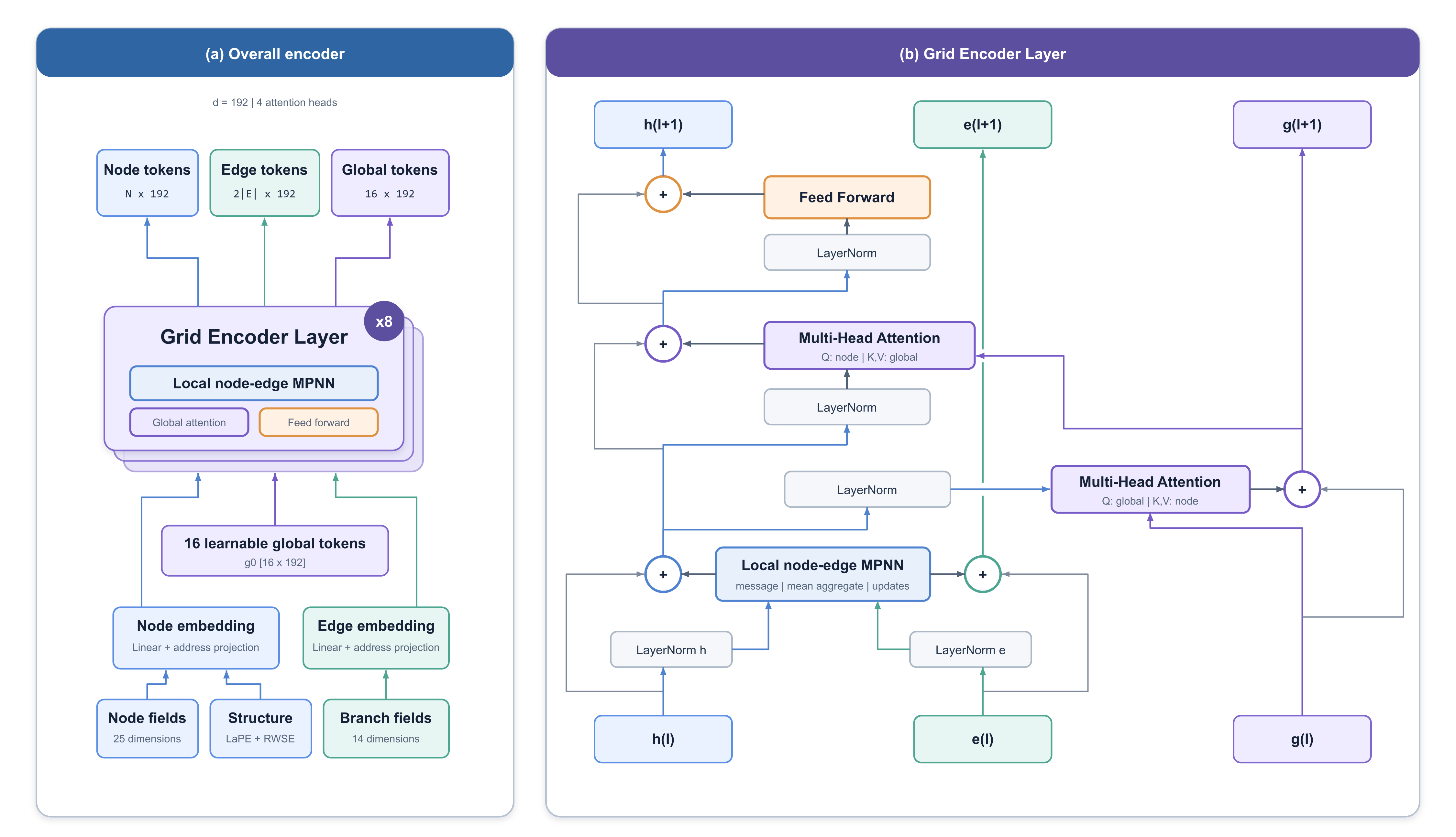}
\caption*{Fig. 2 | Architecture of the task-agnostic grid encoder. (a) Bus and branch inputs are embedded as node and directed-edge tokens and processed with 16 learnable global tokens through eight encoder layers (hidden dimension, 192; attention heads, four). A final layer normalization is applied to the node representations before output. (b) Each layer combines local node-edge message passing, global-token attention and a feed-forward block. Global tokens aggregate node information, which is then read back by the nodes to obtain system-wide context. Layer normalization and residual connections are shown within the layer.}
\end{figure}

The complete set of fields that can be represented by the encoder is summarized in Table 1. The fields supplied to the encoder are determined by the information available for the corresponding computational problem. In particular, for power-flow calculation, quantities that are unknown before solution, including the corresponding bus-voltage variables, are removed from the encoder input together with all branch-flow quantities, as specified in Table 2 and Section 2.3.1. As detailed in Table 1, each node contributes an 81-dimensional input vector and each branch a 22-dimensional input vector before projection into the shared 192-dimensional latent space.

\begingroup\scriptsize
\noindent Table 1 | Encoder input schema: every variable, its physical meaning, unit, normalization and dimensionality, in encoding order (* = p.u. on the 100-MVA system base). The 25 node feature columns form four maskable groups (topology / static / dynamic / operating state); each node and branch record additionally carries per-group missing and masked indicator bits, and masked or missing fields are zeroed in the normalized feature space rather than passed as extra input channels. Normalization statistics are estimated on the encoder-training corpus only.\par\vspace{2pt}
\begin{xltabular}{\textwidth}{>{\RaggedRight}Xl>{\RaggedRight}Xl>{\RaggedRight}Xl}
\toprule
\textbf{Block (encoding order)} & \textbf{Variable(s)} & \textbf{Physical meaning} & \textbf{Unit} & \textbf{Normalization} & \textbf{Dim.} \\
\midrule
\endhead
Node \newline topology & node\_type & Bus type in the AC formulation, one-hot over \{PQ, PV, slack\} & --- & one-hot & 3 \\
Node \newline topology & is\_null & Placeholder bus (no physical bus behind the record) & --- & 0/1 & 1 \\
Node \newline topology & has\_gen; has\_load & Generator / load attached at the bus & --- & 0/1 & 2 \\
Node \newline topology & vbase & Bus nominal (base) voltage & kV & z-score & 1 \\
Node \newline static & vmax; vmin & Bus voltage limits & p.u. & z-score & 2 \\
Node \newline static & pmax; pmin; qmax; qmin & Active- and reactive-power limits of the generation at the bus & p.u.* & z-score & 4 \\
Node \newline dynamic & tj & Rotor inertia constant of the generator & s & signed log, z-score & 1 \\
Node \newline dynamic & xdp & Generator transient reactance X$'$d & p.u. & z-score & 1 \\
Node \newline dynamic & sh; ph & Rated apparent / active power of the generator & p.u.* & z-score & 2 \\
Node \newline operating state & pg; qg & Solved active / reactive generation at the bus & p.u.* & z-score & 2 \\
Node \newline operating state & pl; ql & Active / reactive load demand at the bus & p.u.* & z-score & 2 \\
Node \newline operating state & v & Solved bus-voltage magnitude & p.u. & z-score & 1 \\
Node \newline operating state & theta & Solved bus-voltage phase angle & rad & raw (radians) & 1 \\
Node \newline operating state & v\_set & Generator terminal-voltage set point & p.u. & z-score & 1 \\
Node \newline metadata & adjustable & Bus hosts a dispatchable unit (addressing metadata) & --- & 0/1 & 1 \\
Node \newline structural (appended) & LaPE + validity flags & Laplacian positional encoding of the topology, with per-component validity flags & --- & --- & 16 + 16 \\
Node \newline structural (appended) & RWSE & Random-walk structural encoding (return probabilities over 16 steps) & --- & --- & 16 \\
Node \newline identity (appended) & node\_random & Sample-local random identity channel (within-case addressing cue only) & --- & --- & 8 \\
Node input $\rightarrow$ node token & (all of the above) & Concatenated node input vector, linearly projected into the shared latent space & --- & --- & 81 $\rightarrow$ 192 \\
Branch \newline topology & branch\_type & Branch type, one-hot over \{AC line, series compensation, two-winding transformer, phase shifter, three-winding-transformer equivalent, shunt compensation\} & --- & one-hot & 6 \\
Branch \newline static & r; x; b\_half & Series resistance and reactance; half charging susceptance & p.u. & r: signed log, z; x, b\_half: z-score & 3 \\
Branch \newline operating state & mark & In-service status of the branch & --- & raw 0/1 & 1 \\
Branch \newline operating state & pi; qi; pj; qj & Solved active / reactive power flow at the two branch terminals & p.u.* & z-score & 4 \\
Branch \newline identity (appended) & branch\_random & Loop-identity random channel, shared by the two directed edges of one physical branch & --- & --- & 8 \\
Branch input $\rightarrow$ branch token & (all of the above) & Concatenated branch input vector, projected into a directed-edge token (each physical branch appears as two directed edges with the same features) & --- & --- & 22 $\rightarrow$ 192 \\
Global tokens & g0 & Sixteen learnable global tokens initialized per graph (not derived from input fields) & --- & --- & 16 $\times$ 192 \\
\bottomrule
\end{xltabular}
\endgroup

The encoder was pretrained without downstream task supervision using masked reconstruction of grid attributes. For each power-system sample, the input features were first normalized and then randomly masked at the level of predefined node-feature groups, including topology, static equipment parameters, available dynamic parameters and operating-state variables. Masked fields retained their original input dimensions but were set to zero in the normalized feature space before being passed to the encoder. The resulting node representations were supplied to reconstruction heads that were trained jointly with the encoder to recover the original masked attributes. Reconstruction loss was evaluated only at entries that were both masked and valid in the source data. Continuous variables were trained using Huber loss, whereas categorical and Boolean variables used binary cross-entropy. The operating-state group was assigned twice the reconstruction weight of the other feature groups. Auxiliary reconstruction heads were trained jointly with the encoder to recover the sample-local node and branch identity channels using a mean-squared-error term with a weight of 0.02. This auxiliary objective encouraged preservation of element addressability, while the identity channels themselves were used only as within-case addressing cues rather than as physical grid variables.

Encoder pretraining used 6,866 operating cases from 67 topology families, comprising CEPRI36, IEEE39, IEEE118 and 64 synthetic transmission-network families. Eight additional synthetic families were used for validation and 12 larger synthetic families were reserved for representation-level transfer evaluation; their construction and operating-condition sampling are described in Section 2.4. Normalization statistics were estimated exclusively from the encoder-training corpus and were fixed after pretraining. The reference encoder was optimized for 6,000 steps using AdamW with a learning rate of 3$\times$10\textsuperscript{-4}, weight decay of 0.01, batch size 16, unit-norm gradient clipping, a 5\% warm-up period and cosine learning-rate decay. Checkpoint selection was based on the masked field-reconstruction objective and excluded the auxiliary identity-reconstruction term.

After self-supervised pretraining, the encoder parameters and normalization statistics were fixed for all downstream computations. The resulting frozen encoder \textit{E}\textit{\textsubscript{$\phi$}}\textit{\textsubscript{* }}produced a case-dependent representation \textit{Z}\textit{\textsubscript{i}} = \textit{E}\textit{\textsubscript{$\phi$}}\textit{\textsubscript{*}}(\textit{G}\textit{\textsubscript{i}}), where \textit{$\phi$}\textit{\textsuperscript{*}}denotes the frozen encoder parameters. The representation changes with network structure and operating condition, but the mapping used to obtain it remains unchanged across downstream tasks. Task-specific quantities, including specified power-flow information, voltage-violation descriptors, requested operating conditions and fault information, are introduced only after grid encoding and are supplied to the corresponding task-specific reasoners described in Section 2.3.

\subsection*{2.3 Task-specific computation and physical verification}

Four downstream computations read the same frozen grid representation, and all task-specific information enters only after encoding, as a small set of task-side inputs supplied to the corresponding reasoner. All reasoner outputs were verified independently using PSASP. Task-side inputs of the four reasoners are summarized in Table 2.

\begin{table}[htbp]
\centering
\footnotesize
\caption*{Table 2 | Task-side inputs of the four reasoners. Every reasoner reads the same frozen node, branch and global tokens of the grid representation; the table lists the only additional inputs, injected after encoding, and the information each provides. All outputs are verified independently, either by PSASP replay.}
\begin{tabularx}{\textwidth}{l>{\RaggedRight}X>{\RaggedRight}X}
\toprule
\textbf{Reasoner} & \textbf{Additional input beyond the frozen representation} & \textbf{Information provided} \\
\midrule
Power flow & Specified electrical quantities, supplied as raw values with known-quantity flags & Formulates the problem by specifying which quantities are given and which are to be recovered \\
 & Training-free DC-angle and Q--V estimates & Physics-based initial estimates that involve no learned parameters \\
Reactive-power adjustment & Per-bus low- and over-voltage violation depths & Identifies the location and severity of the violations to be corrected \\
 & Switchable-shunt candidate mask & Delimits the admissible control actions \\
Operating-condition generation & Eight-dimensional request vector: target aggregate-load level and online non-slack generator count & Specifies the operating condition to be synthesized in the latent space \\
Transient-stability assessment & Fault descriptor: faulted branch, fault position, inception time, clearing time and duration & Characterizes the disturbance; the frozen representation encodes only the prefault grid \\
\bottomrule
\end{tabularx}
\end{table}

\subsubsection*{2.3.1 Power-flow calculation}

Power-flow calculation was formulated as recovery of the unknown AC operating state from an unsolved network case. The encoder input retained the network topology, equipment parameters and quantities specified by the corresponding bus types, while unknown bus variables and all branch-flow quantities were removed. The specified quantities were additionally supplied to the downstream reasoner as task-specific inputs with known-value flags. The frozen encoder therefore described the available physical state of the grid, while information required specifically by the power-flow formulation was introduced only after encoding.

As shown in Fig. 3, the power-flow reasoner combines the frozen grid representation with deterministic physical estimates and iterative physical correction. The reasoner contains approximately 2.5 million trainable parameters and reads the frozen node, branch and global tokens through 16 learnable query tokens, two cross-attention layers and four message-passing layers. It predicts branch voltage-angle differences, which are converted to bus phase angles through least-squares potential reconstruction with periodic wrapping, and directly estimates voltage magnitudes and unknown generator active and reactive powers. A zero-training hint channel is supplied as additional task-side information: a loss-aware DC estimate of bus phase angles and a \textit{Q--V} based estimate of voltage magnitudes. These estimates were omitted for 20\% of the training samples so that the reasoner did not depend exclusively on the physical estimates.

The predicted state is subsequently refined by four jointly trained physical-correction rounds. In each round, bus phase angles are updated from the active-power mismatch using a decoupled Newton step based on the B$'$ matrix, while voltage magnitudes are updated from the reactive-power mismatch using the corresponding decoupled voltage step. Both corrections are computed from the network admittances and specified power injections, with a learnable step size for each round that converged to approximately 1.1 after training. The training loss is evaluated on the corrected state, while an auxiliary \textit{L}\textsubscript{1} loss with unit weight is applied to the state predicted before correction. The correction rounds were disabled for 20\% of the training samples. A correction round was skipped when the maximum angle update exceeded 1 rad or the maximum voltage-magnitude update exceeded 0.5 p.u., and the correction procedure was not applied when specified quantities were available for less than 99\% of buses. Each correction round requires sparse linear solves equivalent to one iteration of a fast-decoupled power-flow calculation.

Training used an \textit{L}\textsubscript{1} loss on the unknown electrical quantities, with a weight of 4 for phase angle, together with an \textit{L}\textsubscript{1} Kirchhoff-mismatch loss of unit weight evaluated on the predicted operating state. The reasoner was trained for 2,500 optimization steps using AdamW with a learning rate of 3$\times$10\textsuperscript{-4} and a batch size of 8. The training set contained 9,646 operating cases, including 9,070 cases from the synthetic families and benchmark systems and 576 real-scale cases, with the latter sampled at five times their nominal weight during training. Three independent reasoner realizations were trained for the reported evaluation and accuracy was evaluated on CPU.

\begin{figure}[htbp]
\centering
\includegraphics[width=\linewidth]{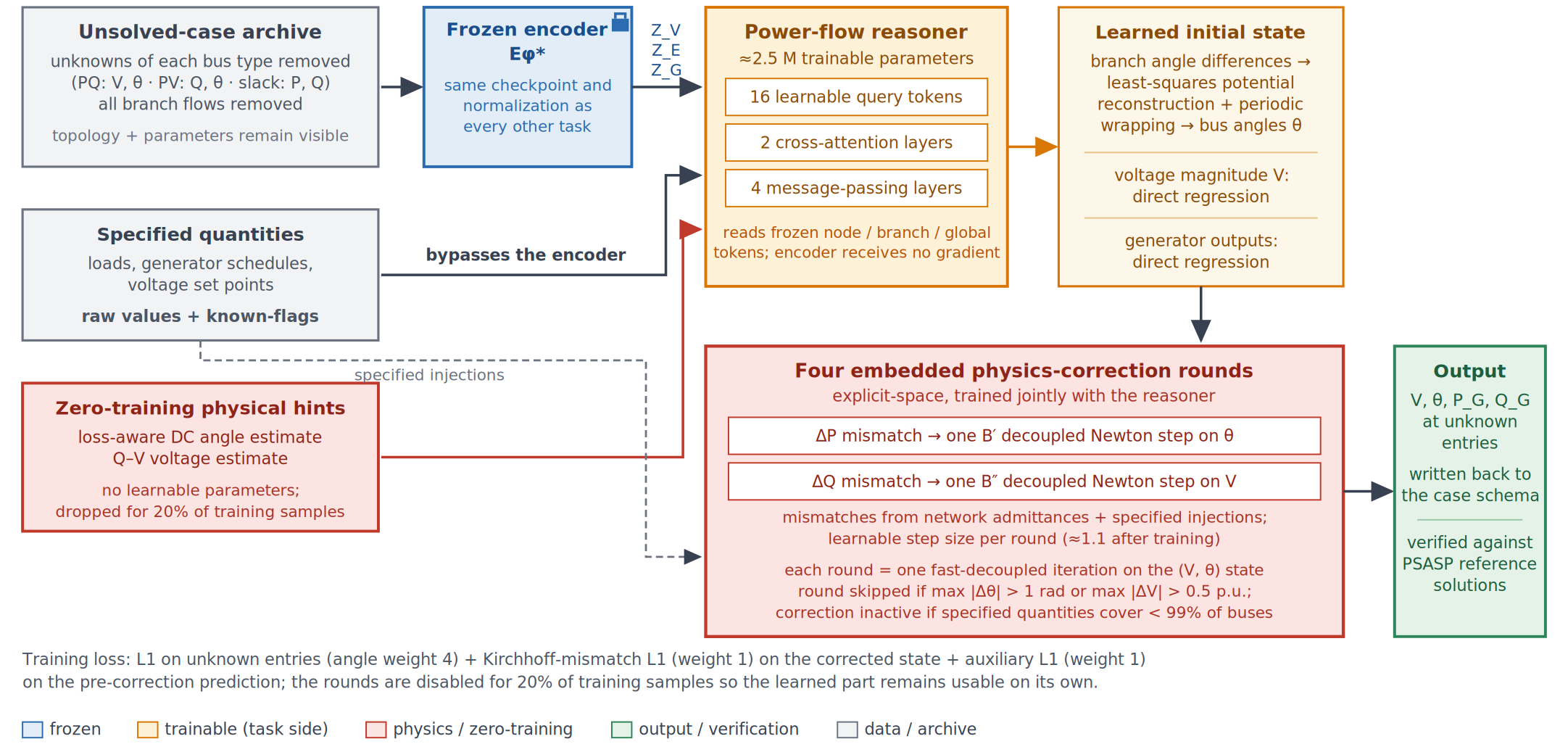}
\caption*{Fig. 3 | Power-flow reasoner with embedded explicit-space physics correction. The unsolved-case archive removes the unknown quantities of each bus type and all branch flows; specified quantities (loads, generator schedules, voltage set points) bypass the encoder and reach the reasoner as raw values with known-flags, and a zero-training hint channel supplies a loss-aware DC angle estimate and a Q-V voltage estimate (dropped for 20\% of training samples). The reasoner (about 2.5 million trainable parameters; 16 learnable query tokens, two cross-attention layers, four message-passing layers) predicts branch angle differences, integrated into bus angles by least-squares potential reconstruction with periodic wrapping, and regresses voltage magnitude and generator outputs directly. Four jointly trained physics-correction rounds then apply, per round, one B' decoupled Newton step on the active-power mismatch and one B$''$ step on the reactive-power mismatch in the explicit (\textit{V}, \textit{$\theta$}) state, with a learnable step size per round (about 1.1 after training), divergence guards (1 rad; 0.5 p.u.) and a 99\% coverage gate.}
\end{figure}

\subsubsection*{2.3.2 Reactive-power adjustment}

Reactive-power adjustment was formulated as a corrective-control task for a solved operating state containing voltage violations. The frozen grid representation was combined with task-specific information introduced only after encoding. This information comprised the per-bus undervoltage and overvoltage violation depths, defined relative to the nominal thresholds of 0.93 and 1.08 p.u., respectively, together with a mask identifying switchable-shunt candidates. As shown in Fig. 4, a 16-query, two-layer, four-head reasoner produces a generator-selection logit, a discrete generator voltage-set-point change selected from (-0.05 p.u., -0.02 p.u., 0.02 p.u., 0.05 p.u.), and a shunt-selection logit. The supervised objective combines binary cross-entropy for generator selection, balanced cross-entropy for the voltage-set-point classes and binary cross-entropy for eligible shunt candidates.

At inference, the predicted voltage-set-point and shunt actions are applied to the original operating case and independently evaluated in PSASP; a corrective action is physically acceptable when the power-flow calculation converges and all valid bus voltages lie within the exam band of the grid. For evaluations, the reasoner adds a graph-level count head that selects the number of units to act on and a violation-proximity side feature (graph distance to the nearest violating bus); decoding takes the top-ranked candidate units up to the predicted count. At deployment the learned action is followed, on failure, by a model-guided ladder that enlarges the unit set along the reasoner's ranking (one PSASP call per step) and finally by a predefined engineering rule, after resetting the case to its original state. A locality prescreen restricts the candidate units to those within H hops of the nearest violating bus; it is switched on only when the reasoner's first choice lies at least eight hops farther from the violation than the first choice of the engineering rule, because an offset of four hops or less brings no gain, and H is chosen per grid as the smallest radius for which at most 5\% of cases have no candidate and the median number of candidates is at least eight. The switching threshold and these census parameters were fixed beforehand, in deployment-calibration experiments on earlier violation exams, and were applied unchanged to the paired exams reported here; the radii used for the reported evaluations are the census outputs for those exams.

\begin{figure}[htbp]
\centering
\includegraphics[width=\linewidth]{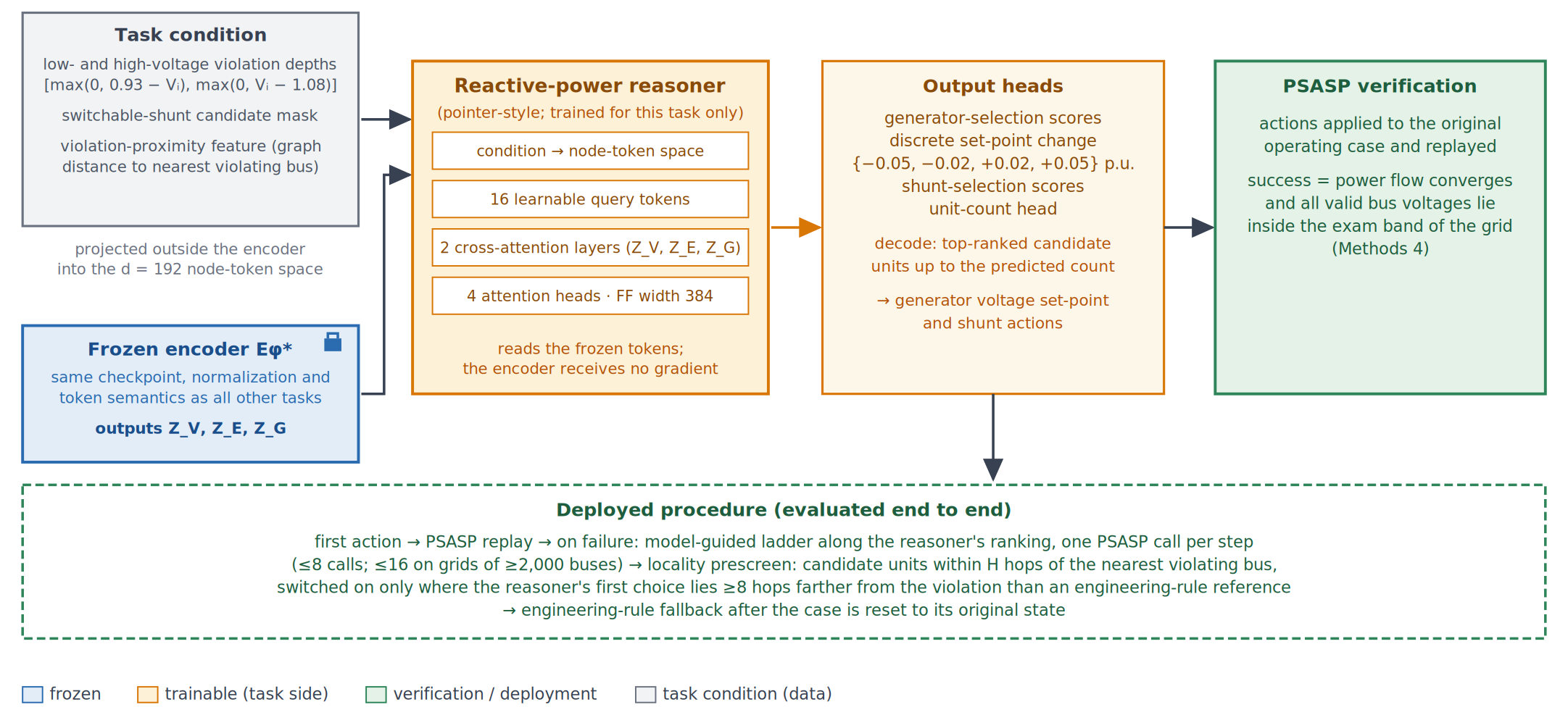}
\caption*{\textbf{Fig. 4 | Reactive-power adjustment reasoner and deployed procedure. The task condition}\textbf{, }\textbf{low- and high-voltage violation depths, the switchable-shunt candidate mask and a violation-proximity feature}\textbf{, }\textbf{is projected outside the encoder into the node-token space and read, together with the frozen tokens, by a pointer-style reasoner (16 learnable query tokens, two cross-attention layers, four attention heads, feed-forward width 384). Its heads return generator selection, a discrete set-point class \{-0.05, -0.02, +0.02, +0.05\} p.u., shunt selection and a unit count; decoding takes the top-ranked candidate units up to the predicted count. The deployed procedure verifies each action in PSASP and, on failure, applies a model-guided ladder along the reasoner's ranking, a locality prescreen and}\textbf{ }\textbf{an engineering-rule fallback (}\textbf{Section 2.3.2}\textbf{)}\textbf{.}}
\end{figure}

\subsubsection*{2.3.3 Operating-condition generation}

Operating-condition generation was formulated as a conditional generation task in which the frozen grid representation was used to construct physically feasible operating states for specified system-level conditions. Operating-condition generation differs from other tasks and uses a conditional generative reasoner. As shown in Fig. 5, a task-side latent denoising-diffusion/readout stack operates on the frozen \textit{d}=192 grid representation together with an eight-dimensional request vector encoding the requested load level and online-unit condition. The denoiser contains eight AdaLN-modulated message-passing layers (9.8 million parameters) and is trained as a T=1000 cosine-noise diffusion model with AdamW, exponential moving-average weights and condition dropout on a corpus of 8,133 cases from 81 families (7,319 for training and 814 for validation) that includes 250 cases of the 3,120-bus and 300 cases of the 2,000-bus system, up-sampled eightfold; the reported checkpoint is the best-validation checkpoint at 27,000 of a planned 90,000 steps. A separate two-layer message-passing readout, retrained on the same corpus with latent-noise augmentation, maps generated latent node representations back to bus loads and generator active outputs; generator voltage set points are taken from the base case. At inference, deterministic 250-step DDIM sampling with classifier-free guidance generates up to eight seeded candidate operating states; the requested aggregate load is imposed exactly on each candidate by rescaling. On the 2,000-bus system and the two synthetic evaluation sets, generator outputs were additionally blended with the base-case outputs with weight 0.5 before simulation; the 3,120-bus evaluations used the generated outputs without base-case blending. An independently fixed convergence prescreener orders the candidates before PSASP evaluation, and the first candidate that converges and satisfies the voltage criterion of the grid is delivered. The delivery success rate therefore characterizes the complete proposal--screen--solve pipeline rather than the diffusion network in isolation. Fidelity to the requested load and online-unit conditions is evaluated separately from physical feasibility.

\begin{figure}[htbp]
\centering
\includegraphics[width=\linewidth]{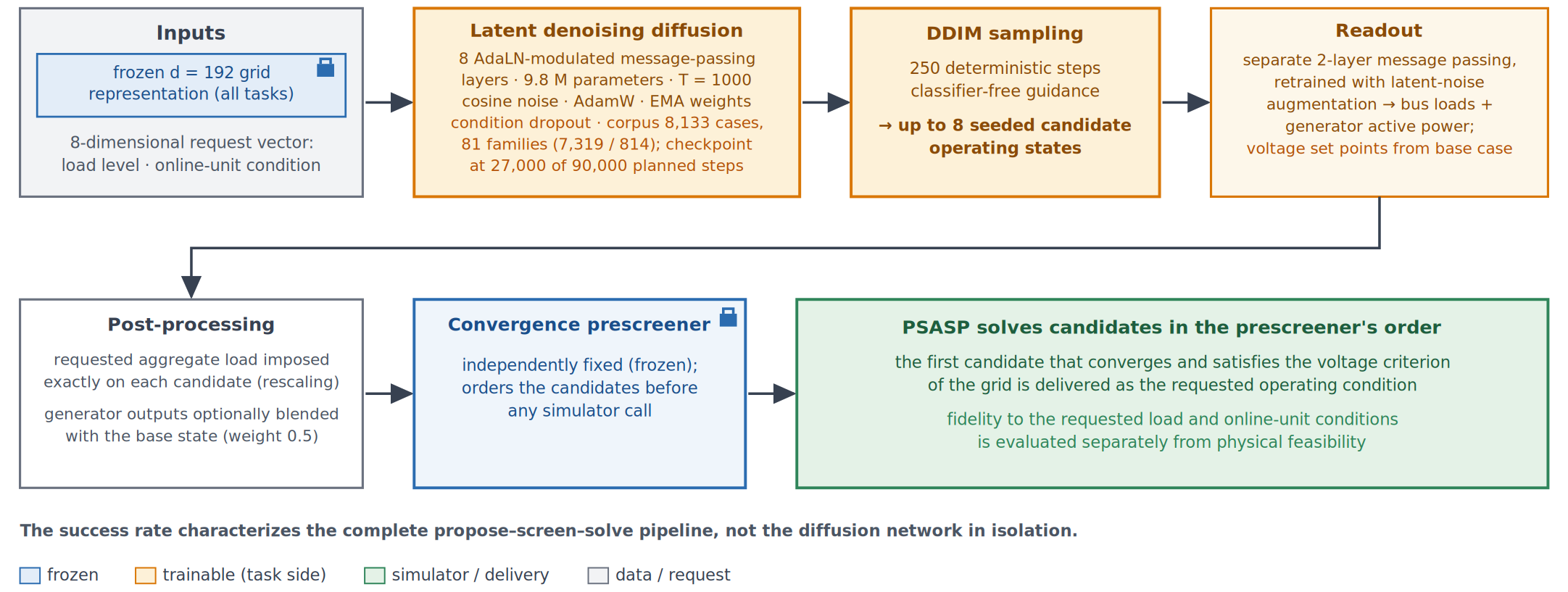}
\caption*{Fig. 5 | Operating-condition generator as a propose-screen-solve pipeline. The frozen d = 192 grid representation and an eight-dimensional request vector (load level, online-unit condition) condition a latent denoising-diffusion model (eight AdaLN-modulated message-passing layers, 9.8 million parameters, T = 1000 cosine noise; trained with AdamW, EMA weights and condition dropout). At inference, 250-step deterministic DDIM sampling with classifier-free guidance produces up to eight seeded candidates; a separate two-layer message-passing readout, retrained with latent-noise augmentation, maps generated latents back to bus loads and generator active power, with voltage set points taken from the base case. The requested aggregate load is imposed exactly on each candidate by rescaling; generator outputs are blended with the base state with weight 0.5 on the 2,000-bus and synthetic evaluations, and used without blending on the 3,120-bus evaluations. An independently fixed convergence prescreener orders the candidates, and PSASP solves them in that order; the first candidate that converges and satisfies the grid's voltage criterion is delivered. The delivery success rate characterizes the complete propose-screen-solve pipeline, not the diffusion network in isolation.}
\end{figure}

\subsubsection*{2.3.4 Transient-stability assessment}

For transient-stability assessment, the frozen representation describes the solved prefault system, while disturbance information is explicitly provided outside the encoder. The event representation combines the frozen embedding of the faulted branch with four continuous descriptors---fault location K\%, inception time, clearing time and duration---through a two-layer multilayer perceptron. No post-fault trajectory or stability outcome is provided to the encoder or reasoner. As shown in Fig. 6, a 16-query, two-layer, four-head reasoner with feed-forward width 384 predicts an instability logit, five dynamic-security subtype logits, a critical-clearing-time estimate and generator-wise loss-of-synchronism-group scores. The composite objective combines instability classification, unstable-event subtype classification, critical-clearing-time regression and machine-group classification. Five reasoner realizations are trained independently while the encoder remains fixed, under a 40,000-step schedule with early stopping; the five realizations stopped after 12,200--21,000 optimization steps. At inference, the stability decision uses the fixed rule \textit{p} > 0.5, and classification performance is reported as accuracy and as recall of unstable events; the auxiliary heads provide complementary information. Physical reference labels are obtained from electromechanical transient simulation rather than from the learned model. A topology family omitted from reasoner labels but exposed during encoder development is consequently described as held out from reasoner training and encoder-exposed.

\begin{figure}[htbp]
\centering
\includegraphics[width=\linewidth]{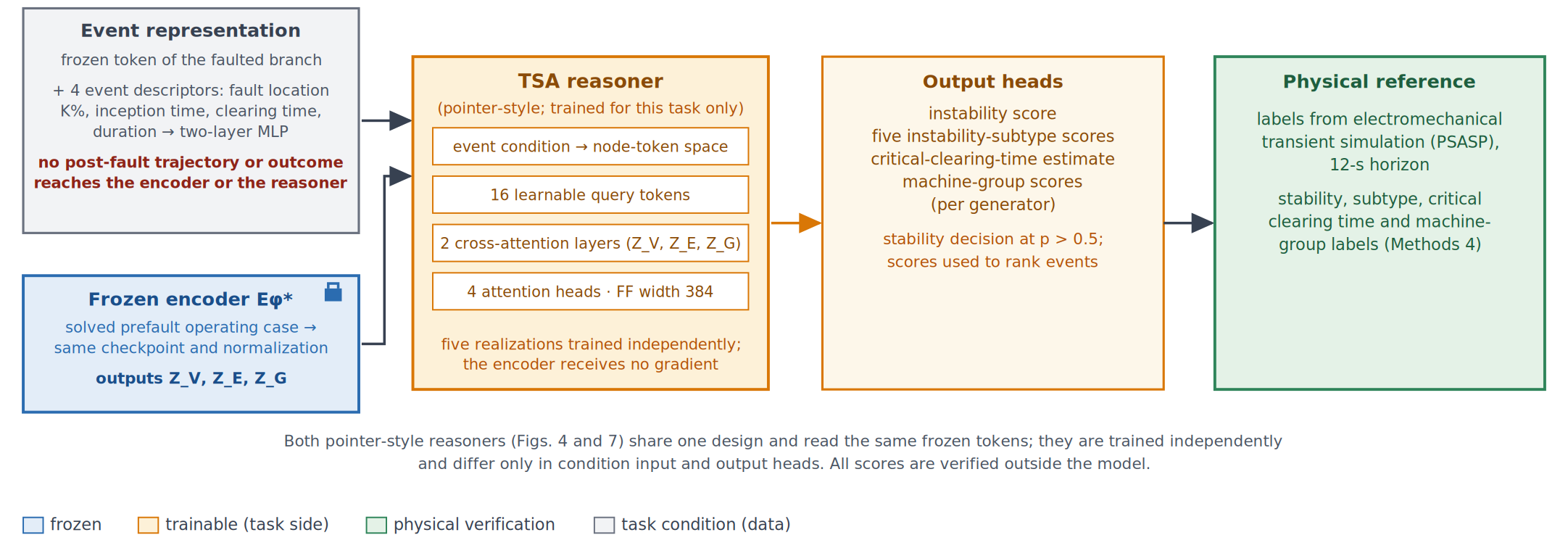}
\caption*{Fig. 6 | Transient-stability assessment reasoner. The event representation combines the frozen token of the faulted branch with four continuous descriptors (fault location K\%, inception time, clearing time and duration) through a two-layer MLP; no post-fault trajectory or outcome reaches the encoder or the reasoner. The same pointer-style core as in Fig. 4 (16 learnable query tokens, two cross-attention layers over the node, branch and global tokens, four attention heads, feed-forward width 384) returns an instability score, five instability-subtype scores, a critical-clearing-time estimate and per-generator machine-group scores; five realizations are trained independently, and all labels come from electromechanical transient simulation.}
\end{figure}

\subsection*{2.4 Power-system datasets}

Steady-state and transient datasets were generated and evaluated using PSASP. The study used benchmark systems including CEPRI36, IEEE39, IEEE118 and IEEE300, independently generated synthetic transmission-network families, and larger transmission systems including case3120sp, ACTIVSg2000, ACTIVSg25k, ACTIVSg70k, PEGASE 1354 and PEGASE 2869. Different datasets were constructed for encoder pretraining, representation evaluation and each downstream computational task. Steady-state samples were obtained by perturbing operating conditions or network configurations and solving the resulting cases in PSASP, whereas transient samples were generated through electromechanical time-domain simulation. The corresponding data-generation procedures are described below.

\subsubsection*{2.4.1 Encoder pretraining and representation-evaluation data}

Encoder pretraining used 6,866 operating cases from 67 topology families, comprising 64 synthetic transmission-network families together with CEPRI36, IEEE39 and IEEE118. For each synthetic training family, operating conditions were generated by varying the system load, generator dispatch, generator availability and terminal-voltage set points around the corresponding base case. The global load level was sampled from 0.95, 1.05 and 1.15 times the base load, with an additional independent variation of $\pm$3\% applied to individual loads. Generator active-power dispatch was redistributed according to the base-case dispatch with random perturbations of $\pm$15\%, subject to generator limits, and different generator-availability patterns were also sampled. Generator voltage set points were varied within $\pm$0.03 p.u. Cases were solved in PSASP and converged operating points with a minimum bus voltage of at least 0.85 p.u. were retained.

Eight additional synthetic families were generated using the same procedure for validation. Representation transfer to larger synthetic networks was evaluated on 12 independently generated families at 400, 700 and 1,000 buses. To avoid excessive infeasibility in these larger systems, their operating conditions were sampled over narrower load and voltage-setpoint ranges and generator decommitment was not applied. Representation evaluation additionally included operating cases from the larger transmission systems listed above. All normalization statistics used by the encoder were estimated from the pretraining corpus and fixed thereafter.

Topology sensitivity was evaluated using matched N-0, N-1 and N-2 cases generated after encoder training. For each sampled operating condition, the intact network was compared with cases in which one or two in-service branches were opened while network connectivity was preserved. The same underlying operating-condition draw was used across the three network configurations, and only groups for which all variants converged with acceptable voltages were retained. This produced paired cases in which the principal difference was the imposed network-topology change.

\subsubsection*{2.4.2 Power-flow calculation data}

Power-flow samples were generated by perturbing load demand and generation dispatch around the corresponding base operating conditions and retaining cases that converged in PSASP. The resulting solved operating states provided the reference solutions for constructing the unsolved power-flow inputs described in Section 2.3.1. The training corpus contained 9,646 operating cases drawn from synthetic network families, the benchmark systems and selected larger systems. In addition to the synthetic families used for encoder pretraining, 13 independently generated 400-bus families were included specifically for power-flow reasoner training.

Evaluation cases were generated independently of reasoner training. For example, the IEEE118 evaluation set was resampled after training using new operating conditions with total-load factors between 0.90 and 1.07. Larger-system evaluations used separately retained operating cases from ACTIVSg2000, ACTIVSg25k and ACTIVSg70k.

Sensitivity to topology changes was assessed using paired N-0, N-1 and N-2 operating cases generated after reasoner training. For each newly sampled operating condition, the intact network was solved first, followed by cases in which one and two in-service branches were opened while preserving network connectivity. The same load, dispatch and voltage-set-point realization was used across the three variants. A group was retained only when all variants converged and maintained a minimum bus voltage of 0.85 p.u. All paired cases were used exclusively for evaluation, and each retained case was verified to be absent from every training corpus. The corresponding PSASP solutions served as the physical references for power-flow evaluation.

\subsubsection*{2.4.3 Reactive-power adjustment data}

Reactive-power adjustment samples were constructed from converged operating conditions by deliberately introducing voltage violations through increased reactive-power demand. For most systems, reactive loads within two network hops of a selected location were multiplied by factors between 3 and 6, or reactive demand was increased across the network. For systems with 25,000 buses or more, factors between 5 and 10 were used. Cases were retained only when the modified operating condition remained convergent in PSASP and at least one bus voltage violated the examination band. The prescribed voltage band is grid-specific and serves both as the success criterion and as the reference band for the violation-depth input features. The lower limit is 0.93 p.u., relaxed to 0.92 p.u. on IEEE300, whose base case reaches a minimum of 0.929 p.u.; the upper limit is 1.08 p.u. unless the source operating-case pool of the grid already exceeds it, in which case it is raised to the pool maximum rounded up to the next 0.01 p.u. (1.09 p.u. for IEEE300, 1.12 p.u. for PEGASE 1354 and case3120sp, 1.11 p.u. for ACTIVSg25k, and 1.15 p.u. for PEGASE 2869 and ACTIVSg70k), with a numerical tolerance of 5$\times$10\textsuperscript{-4} p.u. in the criterion. The same band values were applied unchanged to the N-1 and N-2 variants of each exam.

Topology-change evaluations were constructed by applying the same reactive-load disturbance to matched N-0, N-1 and N-2 network configurations. The three variants therefore shared the same underlying operating condition and reactive-power disturbance and differed only in branch status and the resulting electrical state. A group was retained only when all three variants converged and exhibited a voltage violation within the prescribed examination range. The locality-prescreen radius used by the deployed controller was fixed from the N-0 cases and then kept unchanged for the corresponding N-1 and N-2 variants. The effectiveness of a predicted corrective action was evaluated independently by applying the action to the original case and resolving the modified operating condition in PSASP.

\subsubsection*{2.4.4 Operating-condition generation data}

The operating-condition generation corpus was constructed by varying total load around the corresponding base cases, redistributing generation in proportion to the required load level and retaining operating states that converged in PSASP. The resulting dataset contained 8,133 operating cases from 81 topology families, of which 7,319 were used for training and 814 for validation. Larger-system samples from ACTIVSg2000 and case3120sp were also included in the training corpus.

Independent request sets were then constructed to evaluate whether the model could generate physically feasible operating states under specified system-level conditions. Each request specified a target aggregate-load level and a target number of online non-slack generators. For case3120sp, 50 requests were generated with load factors between 0.90 and 1.07, and for ACTIVSg2000, 60 requests covered load factors between 0.90 and 1.03. Additional request sets were constructed from the synthetic validation and larger transfer families. Generated candidates were written back to the corresponding power-system case and solved in PSASP, with convergence and voltage compliance used to determine physical feasibility.

For topology-change evaluation, the same requested operating conditions were applied to N-0, N-1 and N-2 versions of each base network. The branch outages were introduced before operating-condition generation, while the requested load level and online-generator condition were kept unchanged. This allowed the effect of network configuration to be evaluated under matched generation requests.

\subsubsection*{2.4.5 Transient-stability assessment data}

Transient-stability samples were generated from solved prefault operating conditions of CEPRI36, IEEE39 and IEEE118. Prefault cases with a minimum bus voltage below 0.905 p.u. were excluded. Metallic three-phase-to-ground faults were then applied at predefined locations on transmission lines, including line-terminal and selected intermediate locations, and different fault-clearing times were used to obtain both stable and unstable responses. Each event was simulated for 12 s in PSASP to obtain bus-voltage, system-frequency and generator-angle trajectories.

An event was classified as unstable if the relative generator-angle separation reached 500$^{\circ}$; if any generator terminal voltage fell below 0.8 p.u. later than 10 s after fault inception, if the minimum bus voltage over the final second of the simulation was below 0.9 p.u., or if any bus voltage exceeded 1.2, 1.3 or 1.4 p.u. for longer than 0.02 s, 0.1 s or 0.5 s, respectively; or if the centre-of-inertia frequency remained outside 0.98-1.02 p.u. for at least 0.5 s. Critical clearing time was determined within 0-0.5 s by binary search with a tolerance of 0.01 s, with cases remaining stable at 0.5 s treated as right-censored. Loss-of-synchronism generator groups were additionally identified for angle-instability events.

The reasoner-training dataset contained 23,250 fault events from IEEE118 and CEPRI36. Evaluation on these two systems used 5,700 events generated from prefault operating conditions excluded from reasoner training. A further 10,500 events from IEEE39 were used to evaluate transfer to a grid for which no transient-stability labels were used in reasoner training.

\section*{Results}

\subsection*{3.1 Power grids spanning diverse network structures and scales}

The experiments covered structurally diverse power systems spanning 36 to 70,000 buses, including benchmark systems, synthetic topology families and large-scale transmission networks. Encoder pretraining used 67 topology families, comprising CEPRI36, IEEE39, IEEE118 and 64 independently generated synthetic transmission-network families of 150--270 buses. Eight additional synthetic families of the same size range were reserved for validation. Representation transfer beyond the encoder-training distribution was evaluated on 12 independently generated synthetic families, with four each at 400, 700 and 1,000 buses, and on six larger transmission systems spanning 1,354-70,000 buses: PEGASE 1354, ACTIVSg2000, PEGASE 2869, case3120sp, ACTIVSg25k and ACTIVSg70k.

The downstream tasks used different subsets of these systems, creating distinct combinations of encoder and reasoner exposure. Some grids were absent from encoder pretraining but were used to train a task-specific reasoner, whereas others were unseen by both the encoder and the corresponding reasoner. Additional task-specific evaluations also included IEEE300. This separation allows representation-level transfer to be distinguished from downstream task transfer. The exposure of each grid to the encoder and individual downstream reasoners is summarized in Table 3. Representation-level generalization across unseen network structures, system scales and topology changes is evaluated in Section 3.2.2, while task-specific transfer is examined in Section 3.3.

\begin{table}[htbp]
\centering
\footnotesize
\caption*{Table 3 | Power grids used for representation learning and downstream evaluation}
\begin{tabularx}{\textwidth}{>{\RaggedRight}Xl>{\RaggedRight}X>{\RaggedRight}X}
\toprule
\textbf{Grid / family} & \textbf{Buses} & \textbf{Encoder}\textbf{ }\textbf{grid}\textbf{ exposure} & \textbf{Reasoner }\textbf{grid }\textbf{exposure} \\
\midrule
CEPRI36 & 36 & Seen(pretraining) & Power flow calculation (Seen); TSA (Seen) \\
IEEE39 & 39 & Seen(pretraining) & Power flow calculation (Seen); TSA (Unseen) \\
IEEE118 & 118 & Seen(pretraining) & Power flow calculation(Seen); Reactive-power adjustment (Seen); TSA (Seen) \\
Synthetic 150--270-bus families & 150--270 & Seen(64 training families; 8 validation families) & Power flow calculation (Seen); Operating-condition generation (Unseen) \\
Synthetic 400/700/1,000-bus families & 400, 700, 1,000 & Unseen(Generalization evaluation) & Operating-condition generation (Unseen) \\
IEEE300 & 300 & Unseen(not used) & Power flow calculation (Seen); Reactive-power adjustment (Unseen) \\
case3120sp & 3,120 & Unseen (generalization evaluation) & Reactive-power adjustment (Seen); Operating-condition generation (Seen) \\
PEGASE 1354 & 1,354 & Unseen (generalization evaluation) & Reactive-power adjustment(Unseen) \\
PEGASE 2869 & 2,869 & Unseen (generalization evaluation) & Reactive-power adjustment(Unseen) \\
ACTIVSg2000 & 2,000 & Unseen(Generalization evaluation) & Power flow calculation(Seen); Reactive-power adjustment (Seen);  Operating-condition generation (Seen) \\
ACTIVSg25k & 25,000 & Unseen (generalization evaluation) & Power flow calculation (Unseen); Reactive-power adjustment (Unseen) \\
ACTIVSg70k & 70,000 & Unseen (generalization evaluation) & Power flow calculation (Unseen); Reactive-power adjustment (Unseen) \\
\bottomrule
\end{tabularx}
\end{table}

\subsection*{3.2 The unified grid representation generalizes across structures and scales}

\subsubsection*{3.2.1 The representation preserves element addressability and operating-state information}

Power-system computation is inherently associated with specific network elements. Bus voltages must remain linked to their physical buses, control actions must be applied to the correct generators or shunts, and disturbance information must identify the corresponding transmission branches. A reusable grid representation must therefore preserve both the correspondence of individual network elements and the electrical operating information associated with them. The encoder providing this representation was pretrained, self-supervised, on 6,866 operating cases from 67 topology families and then frozen; no downstream task updates it. To assess whether the principal representation-level findings were reproducible rather than specific to a single trained encoder, we evaluated three independently trained encoder realizations, hereafter denoted seed0, seed1 and seed2 according to their initialization seeds.

We first examined whether individual buses remained identifiable after encoding. Across 20 operating cases from eight validation grid families, the same physical system was encoded twice under independently redrawn sample-local addressing cues. For each valid bus, its counterpart in the second representation was identified from the most similar normalized bus representation. The bus-correspondence accuracy was defined as the fraction of buses assigned to the same physical bus and was averaged over the 20 cases.  As shown in Table 4, across the three encoder realizations, correspondence accuracy reached 0.9973--0.9976. Thus, the encoded bus variables remained almost uniquely associated with their physical locations despite changes in the auxiliary addressing realization.

A more stringent test considered parallel branches having identical terminal buses and identical electrical parameters. Four such branch pairs were constructed across four validation families. For each pair, let \textbf{z}\textsubscript{1} and \textbf{z}\textsubscript{2} denote the encoded branch-token representations of the two physically distinct but electrically identical parallel branches. Their separation was quantified by cosine distance \textit{d}\textsubscript{cos}=1-cos(\textbf{z}\textsubscript{1}, \textbf{z}\textsubscript{2}), where \textit{d}\textsubscript{cos}=0 indicates identical representation directions and larger values indicate greater separation in representation space. We used 10\textsuperscript{-3} as the numerical separation criterion. The minimum distance across the four branch pairs was 0.0447-0.0800 for the three encoders, indicating that even the least-separated pair remained distinguishable in representation space.  Thus, the encoder preserved the distinction between separately addressable physical branches when an explicit within-case identity cue was provided, rather than distinguishing physically identical branches from network physics alone.

\begin{table}[htbp]
\centering
\footnotesize
\caption*{Table 4 | Preservation of individual grid elements across encoder realizations}
\begin{tabularx}{\textwidth}{>{\RaggedRight}Xlll}
\toprule
\textbf{Evaluation metric} & \textbf{Encoder seed0} & \textbf{Encoder seed}\textbf{1} & \textbf{Encoder seed}\textbf{2} \\
\midrule
Physical-bus correspondence accuracy & 0.997580 & 0.997394 & 0.997335 \\
Minimum separation of electrically identical parallel branches & 0.0447 & 0.0582 & 0.0800 \\
\bottomrule
\end{tabularx}
\end{table}

Preserving the correspondence of physical elements is necessary but not sufficient for power-system computation. The representation must also retain information about the electrical operating state. We therefore examined whether two fundamental steady-state variables, bus-voltage magnitude V and phase angle $\theta$, could be recovered after these quantities were masked from the encoder input. This experiment evaluates the electrical information retained in the frozen representation. Recovery accuracy was evaluated in physical units using mean absolute error (MAE) over all valid masked bus entries.

Both quantities were recovered consistently across the three encoder realizations  as shown in Table 5. Voltage magnitude was recovered with an MAE of 0.00430--0.00432 p.u. and phase angle with an MAE of 0.261--0.287 rad. The frozen representation therefore retains bus-level operating-state information, and retains \textit{V} more precisely than\textit{ $\theta$}; the additional angle precision that power-flow computation requires is supplied on the task side by the physical estimates and correction of Section 3.3.1.

\begin{table}[htbp]
\centering
\footnotesize
\caption*{Table 5 | Recoverability of steady-state electrical variables across encoder realizations}
\begin{tabular}{lllll}
\toprule
\textbf{Electrical quantity} & \textbf{Metric} & \textbf{Encoder seed0} & \textbf{Encoder seed1} & \textbf{Encoder seed2} \\
\midrule
Voltage magnitude \textit{V} & MAE (p.u.) & 0.004317 & 0.004296 & 0.004319 \\
Phase angle \textit{$\theta$} & MAE (rad) & 0.261162 & 0.286920 & 0.262784 \\
\bottomrule
\end{tabular}
\end{table}

\subsubsection*{3.2.2 The representation generalizes across unseen grids and topology changes}

A reusable grid representation should remain informative when applied to network structures beyond those encountered during encoder training. We therefore evaluated the frozen reference encoder (seed0) on grids whose topology families and system scales were absent from encoder training: twelve synthetic families of approximately 400, 700 and 1,000 buses (565 operating cases; 200, 165 and 200 cases in the three size groups) and six real-scale grids of 1,354--70,000 buses: ACTIVSg2000 (20 cases), PEGASE 1354 and PEGASE 2869 (90 cases each), case3120sp (100 cases), ACTIVSg25k (100 cases) and ACTIVSg70k (40 cases). The eight validation families of 150--270 buses (404 cases) served as the source grids. On each grid set we repeated the bus-addressing, parallel-branch and voltage- and phase-angle recovery tests of Section 3.2.1. To test the representation under topology changes within a grid, we additionally probed paired operating cases generated after training: each draw of loads, dispatch and voltage set points was solved with all branches in service (N-0), with one randomly chosen in-service branch opened (N-1) and with a second branch opened (N-2), keeping the network connected (335 source-grid cases, 125--147 cases per synthetic size group , 41 ACTIVSg2000 cases, 56 ACTIVSg25k cases and 22 ACTIVSg70k cases). Because the larger families and all six real-scale grids were excluded from encoder training, these comparisons evaluate encoder-level generalization across  changes in network structure, system scale and topology.

To quantify the change in recovery accuracy, we compared the physical-unit MAE on the unseen larger grids with that on the source grids,

\begin{equation}
\rho = \frac{\mathrm{MAE}_{\mathrm{target}}}{\mathrm{MAE}_{\mathrm{source}}}
\end{equation}

where \textit{$\rho$}\textit{ }=1 indicates unchanged recovery error and values above one indicate a larger recovery error on the unseen grids.

Table 6 summarizes the results by grid size; its source-grid values come from a rerun of the Section 3.2.1 probe in the same computing environment as the other rows and lie within about 1\% of Table 5. Element addressability was preserved throughout: bus-correspondence accuracy stayed above 0.99 on all synthetic grids up to 1,000 buses, reached 0.97 on ACTIVSg2000 and declined with scale to 0.79 on the 70,000-bus grid, against random-match references between 7$\times$10\textsuperscript{-3} and 2$\times$10\textsuperscript{-5}, and electrically identical parallel branches remained separable on every grid on which such pairs exist. Voltage-magnitude information transferred to the larger synthetic grids with little loss: the recovery MAE rose from 0.00435 p.u. on the source grids to 0.00473--0.00474 p.u. at 400 and 700 buses and 0.00557 p.u. at 1,000 buses (\textit{$\rho$}\textit{\textsubscript{V}} = 1.09--1.28). Phase angle was more sensitive to the scale shift, and the increase was concentrated at the largest size: $\rho$$\theta$ was 1.89--1.90 at 400 and 700 buses and 3.05 at 1,000 buses. On the six real-scale grids the voltage-magnitude recovery MAE was 0.0124--0.0307 p.u. (\textit{$\rho$}\textit{\textsubscript{V}} = 2.8--7.1) and the phase-angle MAE 0.157--0.626 rad: the frozen representation carries the identity of the network elements of these grids, while the operating-state information that the downstream tasks require is supplied on the task side by physical estimates and correction (Section 3.3.1). Opening one or two branches left the representation essentially unchanged. The paired cases are new operating draws, so their N-0 values differ from the main columns of Table 6 and the comparison of interest is across the three paired columns: between the N-0, N-1 and N-2 variants, the voltage-recovery MAE changed by at most 0.0002 p.u. and the phase-angle MAE by at most 0.015 rad (Table 6, last three columns), and bus-correspondence accuracy and parallel-branch separation changed by less than 0.001 and 0.0002, respectively.

\begin{table}[htbp]
\centering
\footnotesize
\caption*{Table 6 | Grid-element identity and electrical information retained on unseen larger synthetic grids and real-scale grids of 1,354-70,000 buses (encoder seed0; the last three columns are paired cases with all branches in service and with one and two branches opened)}
\begin{tabularx}{\textwidth}{l>{\RaggedRight}X>{\RaggedRight}X>{\RaggedRight}X>{\RaggedRight}X>{\RaggedRight}X>{\RaggedRight}X>{\RaggedRight}X>{\RaggedRight}X}
\toprule
\textbf{Grid  (buses)} & \textbf{Bus addressing (random reference)} & \textbf{Parallel-branch separation} & \textbf{\textit{V}}\textbf{ MAE} \newline \textbf{(}\textbf{p.u.}\textbf{)} & \textbf{\textit{$\theta$}}\textbf{ MAE} \newline \textbf{(}\textbf{rad}\textbf{)} & \textbf{Error relative to source grids}\textbf{ }\textit{$\rho$}\textbf{, }\textbf{\textit{V / $\theta$}} & \textbf{Paired cases N-0: }\textbf{\textit{V / $\theta$}}\textbf{ MAE} & \textbf{Paired cases N-1: }\textbf{\textit{V / $\theta$}}\textbf{ MAE} & \textbf{Paired cases N-2: }\textbf{\textit{V / $\theta$}}\textbf{ MAE} \\
\midrule
Source grids (150--270) & 0.9979 (7.1$\times$10\textsuperscript{-3}) & 0.0437 & 0.00435 & 0.2639 & 1 / 1 & 0.00443 / 0.2959 & 0.00452 / 0.3090 & 0.00462 / 0.3103 \\
Synthetic grids, 400 & 0.9948 (3.5$\times$10\textsuperscript{-3}) & 0.0570 & 0.00473 & 0.4994 & 1.09 / 1.89 & 0.00437 / 0.4856 & 0.00443 / 0.4936 & 0.00456 / 0.5000 \\
Synthetic grids, 700 & 0.9907 (2.0$\times$10\textsuperscript{-3}) & 0.0048 & 0.00474 & 0.5026 & 1.09 / 1.90 & 0.00394 / 0.7430 & 0.00397 / 0.7443 & 0.00399 / 0.7355 \\
Synthetic grids, 1,000 & 0.9923 (1.4$\times$10\textsuperscript{-3}) & 0.0984 & 0.00557 & 0.8049 & 1.28 / 3.05 & 0.00457 / 0.8559 & 0.00460 / 0.8644 & 0.00462 / 0.8662 \\
PEGASE 1354 & 0.8848 (7.8$\times$10\textsuperscript{-4}) & No such branch pairs & 0.02155 & 0.1566 & 4.96 / 0.59 & --- & --- & --- \\
ACTIVSg2000 & 0.9722 (7.4$\times$10\textsuperscript{-4}) & 0.0013 & 0.01235 & 0.4685 & 2.84 / 1.78 & 0.01268 / 0.4445 & 0.01266 / 0.4452 & 0.01264 / 0.4481 \\
PEGASE 2869 & 0.9028 (3.8$\times$10\textsuperscript{-4}) & 0.0283 & 0.01782 & 0.2856 & 4.10 / 1.08 & --- & --- & --- \\
case3120sp & 0.8998 (4.0$\times$10\textsuperscript{-4}) & 0.0191 & 0.03073 & 0.3083 & 7.07 / 1.17 & --- & --- & --- \\
ACTIVSg25k & 0.8480 (5.8$\times$10\textsuperscript{-5}) & 0.0012 & 0.01434 & 0.3322 & 3.30 / 1.26 & 0.01360 / 0.3254 & 0.01360 / 0.3253 & 0.01360 / 0.3253 \\
ACTIVSg70k & 0.7919 (2.0$\times$10\textsuperscript{-5}) & 0.0018 & 0.01284 & 0.6256 & 2.96 / 2.37 & 0.01247 / 0.6504 & 0.01247 / 0.6501 & 0.01247 / 0.6507 \\
\bottomrule
\end{tabularx}
\end{table}

As a complementary representation-level diagnostic, we also evaluated reconstruction across a broader set of topology, equipment and operating-state fields using a fixed masking pattern. The test covered 164 cases and 268,647 valid masked entries. Because these fields included both continuous numerical variables and discrete grid attributes, continuous fields were evaluated using Huber loss, whereas categorical and Boolean fields were evaluated using binary cross-entropy. Reconstruction losses were first averaged within each predefined field group and then combined using the predefined group weights to obtain an aggregate representation score. The resulting aggregate score was 0.1751 for the reference encoder seed0, with lower values indicating better overall reconstruction. Because this aggregate score combines heterogeneous variables and loss functions, it is used as an overall diagnostic of representation quality rather than as an accuracy measure in physical units.

\subsection*{3.3 A frozen grid representation supports heterogeneous power-system tasks across grid structures}

Power-system analysis and operation require different computations over the same underlying grid. We therefore tested whether one frozen representation could support four heterogeneous power-system tasks: power-flow calculation, reactive-power adjustment, operating-condition generation and TSA---without task-specific modification of the encoder. These tasks span steady-state analysis, corrective control, operating-state generation and dynamic-security assessment, providing distinct tests of whether the same grid representation can be reused across different computational objectives.

Unless otherwise stated, all downstream evaluations in this section used the same frozen reference encoder, corresponding to encoder seed0 in Section 3.2. Each task used an independently trained reasoner that read the resulting frozen representation together with task-specific information, such as voltage-violation descriptors, requested operating conditions or fault information. Thus, task specialization occurred entirely outside the encoder. Because the tasks have different physical objectives, performance was assessed using task-specific functional and physical endpoints rather than a common accuracy metric. All downstream outputs were independently evaluated using the PSASP.

\subsubsection*{3.3.1 Power-flow calculation}

Power-flow calculation evaluates whether the frozen grid representation can support recovery of an AC operating state from an unsolved case. The power-flow reasoner uses two task-specific physical components in addition to the frozen representation. Deterministic physical hints provide a loss-aware DC estimate of bus phase angles and a \textit{Q}--\textit{V} estimate of voltage magnitudes from quantities available before the AC solution is known, while embedded physics-correction rounds subsequently refine the predicted state using active- and reactive-power mismatches. The former provide non-learned physical priors to the reasoner, whereas the latter form a task-specific physical refinement mechanism within the trained reasoner. The construction of the unsolved-case input, the formulations of the two physical components and the reasoner architecture are described in Section 2.3.1.

In addition to transfer across different grids, we tested whether the trained power-flow reasoner remained effective under topology changes within the same grid. Paired operating cases were generated after training for the four evaluation grids: each new draw of loads, dispatch and voltage set points was solved with all branches in service (N-0), with one randomly chosen in-service branch opened (N-1) and with a second branch opened (N-2), keeping the network connected. All paired cases were generated after training and used only for evaluation. The encoder was trained without outage cases; among the reasoner training corpora, only ACTIVSg2000 contains outage samples (35 historical single-outage operating modes), and the corpora of all other grids contain none. This paired design therefore tests sensitivity to local topology perturbations while controlling for changes in operating condition. Because these paired operating cases were generated independently of the principal evaluation sets, their N-0 results are not expected to match the main evaluation results; the relevant comparison is instead among the matched N-0, N-1 and N-2 case.

Power-flow accuracy was evaluated only on quantities that were unknown before solution. MAE was used as the primary measure, with phase-angle \textit{$\theta$}\textit{ }error reported in degrees after periodic wrapping and voltage-magnitude \textit{V }error in per unit at PQ buses. The maximum absolute error over all evaluated buses, cases and reasoner realizations was reported separately as a tail-error diagnostic rather than as a measure of typical accuracy. Reported mean values were averaged over three independently trained power-flow reasoners.

We first tested whether the trained encoder--reasoner pipeline retained power-flow accuracy when applied across different network structures. The evaluation included both training-exposed grids and previously unseen topologies, thereby testing structural generalization after encoder and reasoner training: the reasoner was trained on 9,646 operating cases from 77 synthetic topology families and from test systems and real-scale grids including CEPRI36, IEEE39, IEEE118, IEEE300 and ACTIVSg2000, with 492 further cases for model selection, and was evaluated on 220 cases from four grids. The corresponding results are summarized in Table 7.  Across the evaluated grids, accurate power-flow recovery was retained on both exposed and unseen structures, including the 25,000- and 70,000-bus systems. Mean \textit{$\theta$} MAE ranged from 0.00014$^{\circ}$ to 0.108$^{\circ}$ , while mean \textit{V}\textit{ }MAE ranged from 1.1$\times$ 10\textsuperscript{-7} to 3.8$\times$10\textsuperscript{-6} p.u., including the previously unseen 25,000- and 70,000-bus systems. The paired topology-perturbation tests showed a similarly stable mean error: from N-0 to N-2, mean \textit{$\theta$}  MAE changed only from 0.00014$^{\circ}$ to 0.00017$^{\circ}$ on IEEE118, from 0.0260$^{\circ}$ to 0.0264$^{\circ}$ on ACTIVSg2000, remained at 0.0038$^{\circ}$ on ACTIVSg25k, and changed from 0.205$^{\circ}$ to 0.207$^{\circ}$ on ACTIVSg70k. The paired V errors behaved in the same way: the mean \textit{V} MAE stayed at 1.1$\times$10\textsuperscript{-7} to 3.3$\times$10\textsuperscript{-6} p.u. across N-0, N-1 and N-2 on all four grids. The mean values nevertheless mask a localized tail effect on ACTIVSg25k: the maximum \textit{$\theta$} error increased from 0.072$^{\circ}$ under N-0 to 28.4$^{\circ}$ under N-1 and N-2, and the maximum \textit{V} error from 2.8$\times$10\textsuperscript{-4} to 0.089--0.108 p.u., because in 1 of the 56 N-1 cases and 2 of the 56 N-2 cases the opened branch left one to three adjacent buses only weakly connected to the rest of the network, and the pipeline did not fully track the resulting large local angle shift; the errors at all other buses, and the mean errors, were unchanged. These results show that the trained representation--reasoning pipeline can transfer across substantial changes in network structure and scale.

\begin{table}[htbp]
\centering
\footnotesize
\caption*{Table 7 | Structural generalization of power-flow calculation across grid topologies and scales (three independently trained reasoners, each with four embedded, jointly trained physics-correction rounds; physical estimates and correction rounds active on all evaluation sets; the last three columns report paired cases generated after training with all branches in service and with one and two branches opened; each paired cell gives the $\theta$ error, mean / max in degrees, and below it the V error, mean / max in p.u.)}
\begin{tabularx}{\textwidth}{ll>{\RaggedRight}X>{\RaggedRight}X>{\RaggedRight}X>{\RaggedRight}X>{\RaggedRight}X>{\RaggedRight}X>{\RaggedRight}X}
\toprule
\textbf{Grid} & \textbf{Buses} & \textbf{Encoder}\textbf{ }\textbf{grid}\textbf{ exposure} & \textbf{Reasoner }\textbf{grid }\textbf{exposure} & \textbf{\textit{V}}\textbf{ MAE} \newline \textbf{mean / max (p.u.)} & \textbf{\textit{$\theta$}}\textbf{ MAE} \newline \textbf{mean / max ($^{\circ}$)} & \textbf{Paired cases N-0}\textbf{ \newline }\textbf{$\theta$ MAE mean / max ($^{\circ}$)}\textbf{ \newline }\textbf{V MAE mean / max (p.u.)} & \textbf{Paired cases N-1}\textbf{ \newline }\textbf{$\theta$ MAE mean / max ($^{\circ}$)}\textbf{ \newline }\textbf{V MAE mean / max (p.u.)} & \textbf{Paired cases N-2}\textbf{ \newline }\textbf{$\theta$ MAE mean / max ($^{\circ}$)}\textbf{ \newline }\textbf{V MAE mean / max (p.u.)} \\
\midrule
IEEE-118 & 118 & Seen & Seen & 1.1$\times$10\textsuperscript{-7} / 8.3$\times$10\textsuperscript{-7} & 0.00014 / 0.00080 & 0.00014 / 0.0008 \newline 1.1$\times$10\textsuperscript{-7} / 9.5$\times$10\textsuperscript{-7} & 0.00016 / 0.0027 \newline 1.1$\times$10\textsuperscript{-7} / 2.5$\times$10\textsuperscript{-6} & 0.00017 / 0.0035 \newline 1.2$\times$10\textsuperscript{-7} / 2.5$\times$10\textsuperscript{-6} \\
ACTIVSg-2000 & 2,000 & Unseen & Seen & 4.6$\times$10\textsuperscript{-7} / 0.00016 & 0.0083 / 0.117 & 0.0260 / 1.01 \newline 7.0$\times$10\textsuperscript{-7} / 0.0016 & 0.0262 / 1.01 \newline 7.0$\times$10\textsuperscript{-7} / 0.0016 & 0.0264 / 1.01 \newline 7.1$\times$10\textsuperscript{-7} / 0.0016 \\
ACTIVSg-25k & 25,000 & Unseen & Unseen & 2.5$\times$10\textsuperscript{-6} / 0.00029 & 0.0038 / 0.080 & 0.0038 / 0.072 \newline 2.5$\times$10\textsuperscript{-6} / 0.00028 & 0.0038 / 28.4 \newline 2.6$\times$10\textsuperscript{-6} / 0.089 & 0.0038 / 28.4 \newline 2.6$\times$10\textsuperscript{-6} / 0.108 \\
ACTIVSg-70k & 70,000 & Unseen & Unseen & 3.8$\times$10\textsuperscript{-6} / 0.0197 & 0.108 / 5.079 & 0.205 / 1.35 \newline 3.3$\times$10\textsuperscript{-6} / 0.0058 & 0.207 / 1.35 \newline 3.3$\times$10\textsuperscript{-6} / 0.0058 & 0.207 / 1.26 \newline 3.3$\times$10\textsuperscript{-6} / 0.0055 \\
\bottomrule
\end{tabularx}
\end{table}

We next asked whether the observed accuracy and topology robustness could be explained primarily by the zero-training physical hints supplied to the reasoner. Table 8 compares the standalone loss-aware DC and\textit{ Q--V} estimates with the final predictions of the trained power-flow reasoner. On all four evaluation grids, the zero-training estimates were substantially less accurate than the corresponding trained predictions.  This comparison shows that the hints provide useful task-specific prior information, but do not by themselves account for the accuracy obtained after learned inference. The physical hints therefore provide useful topology-aware priors, but neither their standalone accuracy nor their behavior under branch outages is sufficient to explain the performance of the trained representation--reasoning framework.

\begin{table}[htbp]
\centering
\footnotesize
\caption*{Table  8 | Zero-training physical estimates(hints) and the trained reasoner on the evaluation sets (the hint for \textit{$\theta$} is the loss-aware DC approximation and the hint for \textit{V} the \textit{Q--V} approximation)}
\begin{tabularx}{\textwidth}{l>{\RaggedRight}X>{\RaggedRight}X>{\RaggedRight}X>{\RaggedRight}X}
\toprule
\textbf{Evaluation set} & \textbf{Hint}\textbf{, }\textbf{zero-training }\textbf{references} \newline \textbf{\textit{V}}\textbf{\textit{ }}\textbf{ MAE (p.u.)} & \textbf{Hint}\textbf{, zero-training references } \newline \textbf{\textit{$\theta$}}\textbf{ MAE ($^{\circ}$)} & \textbf{Hint}\textbf{, trained }\textbf{Reasoner} \newline \textbf{\textit{V}}\textbf{ MAE (p.u.)} & \textbf{Hint}\textbf{, trained r}\textbf{easoner}\textbf{ } \newline \textbf{\textit{$\theta$}}\textbf{ MAE ($^{\circ}$)} \\
\midrule
IEEE118 & 0.00035 & 0.354 & 1.1$\times$10\textsuperscript{-7} & 0.00014 \\
ACTIVSg-2000 & 0.00101 & 1.51 & 4.6$\times$10\textsuperscript{-7} & 0.0083 \\
ACTIVSg-25k & 0.00167 & 1.46 & 2.5$\times$10\textsuperscript{-6} & 0.0038 \\
ACTIVSg-70k & 0.00246 & 17.1 & 3.8$\times$10\textsuperscript{-6} & 0.108 \\
\bottomrule
\end{tabularx}
\end{table}

Finally, we isolated the contribution of the embedded physics correction by comparing reasoners trained with and without the correction rounds while retaining the same frozen representation and physical hints. The corresponding ablation results are reported in Table 9. Without correction, the learned representation and reasoner already provide a transferable initial power-flow estimate, but with substantially lower absolute accuracy: for example, mean \textit{$\theta$}\textit{ } MAE decreases from 0.585$^{\circ}$ to 0.0038$^{\circ}$ on ACTIVSg25k and from 13.8$^{\circ}$ to 0.108$^{\circ}$ on ACTIVSg70k when the correction is included. Together with Table 8, this ablation separates the roles of the three components: the physical hints provide an initial prior, the frozen representation and trained reasoner provide a transferable learned estimate, and the embedded physical correction refines that estimate toward high-precision agreement with the AC reference solution.

\begin{table}[htbp]
\centering
\footnotesize
\caption*{Table 9 | Power-flow reasoner with and without the embedded physics-correction rounds: accuracy on the same evaluation cases}
\begin{tabularx}{\textwidth}{l>{\RaggedRight}X>{\RaggedRight}X}
\toprule
\textbf{Evaluation set} & \textbf{\textit{V}}\textbf{ MAE mean (p.u.),} \newline \textbf{without / with rounds} & \textbf{\textit{$\theta$}}\textbf{ MAE mean ($^{\circ}$),} \newline \textbf{without / with rounds} \\
\midrule
IEEE118 & 0.00039 / 1.1$\times$10\textsuperscript{-7} & 0.165 / 0.00014 \\
ACTIVSg-2000 & 0.00070 / 4.6$\times$10\textsuperscript{-7} & 0.281 / 0.0083 \\
ACTIVSg-25k & 0.00141 / 2.5$\times$10\textsuperscript{-6} & 0.585 / 0.0038 \\
ACTIVSg-70k & 0.00232 / 3.8$\times$10\textsuperscript{-6} & 13.8 / 0.108 \\
\bottomrule
\end{tabularx}
\end{table}

The purpose of the power-flow reasoner in this framework is not to replace a mature numerical power-flow solver. Rather, it establishes a representation-compatible route from an unsolved operating case to the solved-state variables required by subsequent power-system computations. Unlike an external numerical solver, the reasoner reads the same frozen grid representation used by the other downstream tasks and returns the corresponding voltage, angle and generator-state fields through the same representation-based computational interface. This capability is intended to support future end-to-end composition of multiple grid-analysis and control tasks without repeatedly leaving and reconstructing the shared representation. PSASP is therefore retained as the independent AC reference and physical verifier, while the power-flow reasoner provides the representation-reading state-recovery module needed for such a compositional framework.

\subsubsection*{3.3.2 Reactive-power adjustment}

Reactive-power adjustment tests whether low- or high-voltage violations can be corrected through generator voltage-set-point and switchable-shunt actions. The reasoner read the frozen grid representation together with the per-bus low- and high-voltage violation depths, max(0, 0.93-\textit{V}\textit{\textsubscript{i}}) and max(0, \textit{V}\textit{\textsubscript{i}}-1.08), and the set of switchable shunts, and returned the corresponding corrective actions; the predicted actions were applied to the original operating case and independently evaluated in PSASP. The reasoner was trained on 321 reference corrective actions from IEEE118, ACTIVSg2000 and case3120sp (253 for parameter updates and 68 for model selection). It was evaluated on paired violation exams built after training for eight grids (Table 10): for each grid, 25--30 draws of loads, dispatch and voltage set points (28 on ACTIVSg25k and ACTIVSg70k) were each solved with all branches in service (N-0), with one randomly chosen in-service branch opened (N-1) and with a second branch opened (N-2); the same reactive-load amplification was applied to the three variants, and a draw was kept only if all three converged in PSASP with a violation inside the depth limit of the exam. The three variants of a case therefore share one disturbance and differ only in topology, and none of these cases appears in any training corpus. A grid marked Seen in Table 10 supplied training labels from other operating cases of that grid; no evaluation case of any grid was used in training.

Because different combinations of generator voltage set points and shunt actions can produce physically acceptable voltage profiles, performance was evaluated by physical task completion rather than by exact reproduction of a reference action: a case counts as corrected when the PSASP power-flow calculation converges and all valid bus voltages lie within the exam band of the grid, with a tolerance of 5$\times$10\textsuperscript{-4} p.u.

The learned procedure works in steps: the reasoner's first action is applied and verified in PSASP; if it fails, the set of adjusted units is enlarged one step at a time along the reasoner's ranking (the ladder), with one PSASP call per step (up to eight calls; up to 16 on grids of 2,000 buses or more), until a converged state within the band is reached. In deployment, cases that the ladder does not correct are handed to an engineering-rule fallback after the operating case is reset to its original state; the results reported here refer to the learned procedure alone.

On the three grids that supplied training labels, the learned procedure corrected every paired IEEE118 case (1.00 under N-0, N-1 and N-2 alike), every case3120sp case with the intact topology (1.00 under N-0) and 0.77 of the ACTIVSg2000 cases under all three topologies (Table 10). On the five grids absent from reasoner training the success rate was 0.57--0.75. The paired design makes the three columns of Table 10 directly comparable within each grid.

Real-scale grids. The same procedure was applied to violation cases constructed on the real-scale grids: local cases amplified the reactive load within two hops of a chosen bus, and network-wide cases amplified the reactive load of the whole grid. Three of these grids also supplied training labels to the reasoner (IEEE118: 143 labels; case3120sp: 80; ACTIVSg2000: 98), and their evaluation cases were drawn from operating points disjoint from those of the labels; IEEE300, PEGASE 1354 and 2869, ACTIVSg25k and ACTIVSg70k were absent from reasoner training. On grids of 2,000 buses or more the extended ladder of up to 16 calls was used. On grids absent from reasoner training, the reasoner's ranking of candidate units remained usable within a local neighbourhood of the violation, whereas its first choice over the whole grid often lay 5--45 hops from the violating bus; a locality prescreen therefore restricts the candidate units to those within a grid-specific number of hops of the nearest violating bus, and it is switched on only where the reasoner's first choice lies much farther from the violation than that of a simple engineering rule.

Success on the grids absent from reasoner training is lower than on the label-supplying grids for reasons that the deployed procedure makes explicit. First, the reasoner's unit ranking transfers reliably only within a local neighbourhood of the violation: on unseen grids its first choice over the whole grid often lay 5--45 hops from the violating bus, so the deployed configuration relies on the locality prescreen, and correction quality then rests on how much of the violation the local candidate units can control. Second, the exams on the large grids are deep local reactive-load amplifications, and a case counts as corrected only when every valid bus voltage returns to the exam band, so a single unreachable bus fails the whole case. Third, whether the available actions can reach the violating buses at all is a property of the network state: on case3120sp the rate fell from 1.00 (N-0) to 0.80 under N-1 and N-2 because six paired cases derive from one source operating point whose N-1 outage, a 110-kV line, leaves the violating buses beyond the reach of the adjustable units---their minimum voltage stays at 0.89--0.91 p.u. under every rung of the ladder, and the second outage adds nothing. Elsewhere, opening one or two branches left the success rate unchanged (IEEE118, ACTIVSg2000, PEGASE 1354, PEGASE 2869, ACTIVSg25k and ACTIVSg70k) and cost a single case of 30 on IEEE300, under N-2 only (0.60, 0.60, 0.57): a randomly opened branch in a grid of 1,354--70,000 buses is almost always electrically remote from the violating buses, and the initial minimum voltage of these paired cases changed by at most 0.0022 p.u. Branch outages therefore leave the success of the learned procedure unchanged wherever the violation remains within reach of set-point control; the exception is a network condition under which the available set-point actions no longer reach the violating buses.

\begin{table}[htbp]
\centering
\footnotesize
\caption*{Table 10 | Reactive-power adjustment across grids: PSASP-verified success rate of the learned procedure in its deployed configuration on paired violation exams generated after training (the three columns are the same draws solved with all branches in service and with one and two branches opened)}
\begin{tabular}{llllll}
\toprule
\textbf{Grid} & \textbf{Encoder exposure} & \textbf{Reasoner exposure} & \textbf{Paired cases N-0} & \textbf{Paired cases N-1} & \textbf{Paired cases N-2} \\
\midrule
IEEE118 & Seen & Seen & 1.00 & 1.00 & 1.00 \\
case3120sp & Unseen & Seen & 1.00 & 0.80 & 0.80 \\
ACTIVSg2000 & Unseen & Seen & 0.77 & 0.77 & 0.77 \\
IEEE300 & Unseen & Unseen & 0.60 & 0.60 & 0.57 \\
PEGASE 1354 & Unseen & Unseen & 0.64 & 0.64 & 0.64 \\
PEGASE 2869 & Unseen & Unseen & 0.67 & 0.67 & 0.67 \\
ACTIVSg25k & Unseen & Unseen & 0.61 & 0.61 & 0.61 \\
ACTIVSg70k & Unseen & Unseen & 0.75 & 0.75 & 0.75 \\
\bottomrule
\end{tabular}
\end{table}

\subsubsection*{3.3.3 Operating-condition generation}

Operating-condition generation tests whether the frozen grid representation can support the construction of physically feasible operating states for specified system-level operating conditions. Unlike the other tasks, whose reasoners read the frozen representation as input, the generator synthesizes new latent node states in the same representation space and only then decodes them into explicit grid quantities. Each request specifies the target aggregate-load level and the requested number of online non-slack generators. The generation model produces candidate operating states containing bus active and reactive loads and generator active-power outputs; generator voltage set points are taken from the base case of the grid, and the requested aggregate load is imposed exactly on each candidate before simulation. Candidates are subsequently evaluated by PSASP. The generation model was trained on 8,133 operating cases from 81 topology families (7,319 for training and 814 for validation) and evaluated on 360 requested operating conditions across the four evaluation sets of Table 11; the requests of all four sets were additionally re-issued under branch outages.

The primary endpoint was the delivery success rate, defined at the level of the requested operating condition rather than the individual generated candidate: a request counts as successful when the pipeline delivers at least one candidate operating state whose PSASP power-flow solution converges and whose bus voltages satisfy the voltage criterion of the grid. For each request, the final generation pipeline produced up to eight candidate operating states (\textit{k} = 8), which were evaluated sequentially after prescreening; a request with several feasible candidates still counts once, and the denominator is always the number of requested conditions. The voltage criterion is grid-specific: for the synthetic families and the 2,000-bus system, all valid bus voltages within 0.85--1.10 p.u.; for the 3,120-bus system, whose base case itself has eight buses above 1.10 p.u., the per-bus voltage limits specified in the case data, with at most 0.1\% of buses (three of 3,120) outside them, a criterion that the base case satisfies with exactly three buses.

The generator delivered 50 of 50 requested conditions on the 3,120-bus Polish system (load factors 0.90--1.07): all 400 generated candidates converged in PSASP and every request had at least one candidate within the voltage limits. On the 2,000-bus ACTIVSg system it delivered 56 of 60 requested conditions (0.933; load factors 0.90--1.03) with 2.45 PSASP calls per delivered request. On the synthetic families, which were excluded from generation-model training but exposed during development, it delivered 150 of 160 requests (0.938; 2.12 PSASP calls per delivered request) across the eight same-size families of 150--270 buses and 74 of 90 (0.822; 4.22 calls) across the six larger families of 400--1,000 buses (Table 11). The delivered 2,000-bus cases are not copies of training cases: their median distance to the nearest training case of the same topology was 27 times the median nearest-neighbour distance within the 300 training cases of that topology and 5.4 times the typical distance between two training cases.

Physical feasibility alone does not establish that the generated operating state reproduces the numerical condition originally requested. We therefore distinguish conditional fidelity from delivery success; fidelity is evaluated among the requests for which a simulator-feasible operating state was delivered. A delivered operating state satisfied the aggregate-load condition when

\begin{equation}
\left|\, \frac{P_{\mathrm{load,actual}}}{P_{\mathrm{load,target}}} - 1 \,\right| \le 0.05
\end{equation}

and satisfied the online-generation condition when

\begin{equation}
\left|\, n_{\mathrm{online,actual}} - n_{\mathrm{online,target}} \,\right| \le 1
\end{equation}

Joint conditional fidelity required both conditions to be satisfied simultaneously. Its denominator was therefore the number of simulator-feasible delivered requests, rather than the total number of requests or the number of generated candidates.

Physical feasibility and conditional fidelity remain distinct requirements. The aggregate-load condition is satisfied by construction, because the requested load level is imposed on each candidate before simulation. The online-generation condition was satisfied within $\pm$1 unit in 15 of 56 delivered 2,000-bus cases (0.27), 0 of 50 delivered 3,120-bus cases, 32 of 150 same-size synthetic cases (0.21) and 61 of 74 cross-size synthetic cases (0.82), with essentially unchanged rates when the same requests were re-issued on the N-1 and N-2 base cases. The pattern follows the generator-unit data of each grid. On the 2,000-bus system, 389 of the 391 non-slack units have a positive minimum output and are therefore always online in the written-back cases, which pins the delivered online count to 389--391 and the deviation to at most +2 units. The synthetic families have no minimum-output floor; there the generated states place positive output on nearly all units, so the requested count is met only when it lies close to the full unit count, as the cross-size requests do. On the 3,120-bus system, 94 of the 247 non-slack units have a zero minimum output and can genuinely be shut down; the delivered states under-provision by 4 to 21 units and none reaches the requested count. Across all four evaluation sets the pipeline therefore controls the aggregate load exactly, whereas the discrete online-unit count is not effectively controlled by the continuous latent generation pipeline.

\begin{table}[htbp]
\centering
\footnotesize
\caption*{Table 11 | Operating-condition generation across grids and synthetic families (the last two columns re-issue the same requests on the base case with one and with two branches opened)}
\begin{tabularx}{\textwidth}{>{\RaggedRight}Xllll>{\RaggedRight}X>{\RaggedRight}X}
\toprule
\textbf{Grid / evaluation set} & \textbf{Buses} & \textbf{Encoder exposure} & \textbf{Reasoner exposure} & \textbf{Delivery success rate} & \textbf{Delivery success rate}\textbf{ \newline }\textbf{N-1 base case} & \textbf{Delivery success rate}\textbf{ \newline }\textbf{N-2 base case} \\
\midrule
case3120sp & 3,120 & Unseen & Seen & 50/50 = 1.000 & 50/50 = 1.000 & 49/50 = 0.980 \\
ACTIVSg2000 & 2,000 & Unseen & Seen & 56/60 = 0.933 & 57/60 = 0.950 & 56/60 = 0.933 \\
Synthetic 150--270-bus families & 150--270 & Seen & Unseen & 150/160 = 0.938 & 155/160 = 0.969 & 152/160 = 0.950 \\
Synthetic 400/700/1,000-bus families & 400--1,000 & Unseen & Unseen & 74/90 = 0.822 & 74/90 = 0.822 & 73/90 = 0.811 \\
\bottomrule
\end{tabularx}
\end{table}

The 3,120-bus and 2,000-bus evaluations use different voltage criteria (stated above) because the fixed 0.85--1.10 p.u. band cannot be satisfied by the 3,120-bus base case itself; the per-bus criterion is the one under which the training cases of that grid were accepted.

Topology changes within a grid. The requested conditions of the two real-scale sets of Table 11 were re-issued on outage base cases: for each request, the base case with one randomly chosen in-service branch opened (N-1) and with a second branch opened (N-2), keeping the network connected, replaced the intact base case both as the conditioning input of the generator, which reads the outage through the branch-status feature of the frozen representation, and as the case into which the candidates are written back and verified by PSASP; the requests, candidate seeds and delivery protocol were unchanged, so the delivery success rate column is the all-branches-in-service reference for the same requests. On case3120sp the generator delivered 50 of 50 requests on the N-1 base cases and 49 of 50 on the N-2 base cases; all 400 candidates converged in PSASP in each variant, and 395 and 387 of them satisfied the per-bus voltage limits, against 396 with all branches in service. The request lost under N-2 is one whose second outage, a 110-kV line, raised four buses within two hops of the opened line above their 1.12 p.u. limit in every candidate, one bus more than the criterion admits. On ACTIVSg2000 the generator delivered 57 of 60 requests on the N-1 base cases and 56 of 60 on the N-2 base cases, against 56 of 60 with all branches in service, with 2.54 and 2.73 PSASP calls per delivered request (2.45 with all branches in service). Two requests failed on all three topologies; the other failures (two under N-0, one under N-1 and two under N-2) were requests whose eight candidates all failed PSASP verification on that topology while at least one candidate was delivered on another, so the outages changed which marginal requests were delivered rather than how many. The two synthetic evaluation sets were re-issued under the same protocol, with one paired random outage drawn per request: the same-size families delivered 155 of 160 requests on the N-1 base cases and 152 of 160 on the N-2 base cases, against 150 of 160 with all branches in service, and the cross-size families delivered 74 of 90 and 73 of 90, against 74 of 90. Branch outages in the base case therefore leave the delivery success rate of the generator unchanged to within one request on the two real-scale grids, and within five of 160 and one of 90 requests on the synthetic sets, with no monotonic decrease from N-0 to N-2.

\subsubsection*{3.3.4 Transient-stability assessment}

Transient-stability assessment asks whether a frozen prefault representation, combined outside the encoder with the faulted branch, fault position, inception time, clearing time and duration, can classify post-fault instability. The reasoner returned an instability probability and auxiliary predictions for instability subtypes, critical clearing time and the loss-of-synchronism generator group; the positive class was unstable, and the stability decision used the fixed rule \textit{p} > 0.5. The reasoner was trained on 23,250 fault events (operating case $\times$ fault $\times$ clearing duration) from IEEE118 and CEPRI36 and evaluated on 5,700 events from operating cases of the same two grids that were excluded from reasoner training (held out; 3,150 IEEE118 and 2,550 CEPRI36 events) and on 10,500 IEEE39 events (Table 12). Across five realizations, the median accuracy was 0.929 on the held-out IEEE118 events and 0.915 on the held-out CEPRI36 events, with median recall of unstable events of 0.975 and 0.895. Performance on the instability classes is reported as recall and precision on the 3,486 held-out unstable events of the two grids pooled, with the five labelled subtypes merged into three classes---power-angle, voltage and frequency instability (Table 13); an unstable event can belong to several classes, and a class prediction counts as correct when any of its subtypes is predicted. Recall was 88.2--99.2\% and precision 92.9--95.9\% across the three classes.

\begin{table}[htbp]
\centering
\footnotesize
\caption*{Table 12 | Transient-stability assessment across grids}
\begin{tabularx}{\textwidth}{lll>{\RaggedRight}Xl>{\RaggedRight}X}
\toprule
\textbf{Grid} & \textbf{Buses} & \textbf{Encoder exposure} & \textbf{Reasoner exposure} & \textbf{Accuracy } & \textbf{Recall of unstable events} \\
\midrule
IEEE118 & 118 & Seen & Seen (grid in training; evaluation cases excluded) & 0.929 [0.918--0.951] & 0.975 [0.965--0.987] \\
CEPRI36 & 36 & Seen & Seen (grid in training; evaluation cases excluded) & 0.915 [0.904--0.943] & 0.895 [0.878--0.936] \\
IEEE39 & 39 & Seen & Unseen (no training labels from this grid) & 0.678 [0.664--0.696] & 1.000 [0.999--1.000] \\
\bottomrule
\end{tabularx}
\end{table}

IEEE118 and CEPRI36 values are medians [min--max] across five reasoner realizations; IEEE39 values are means [min--max] across the same five realizations. The IEEE39 recall reflects the near-uniform tendency to predict instability on that grid.

\begin{table}[htbp]
\centering
\footnotesize
\caption*{Table 13 | Recall and precision per instability class on the held-out unstable events (IEEE118 and CEPRI36 pooled, 3,486 events; five-realization median [min--max]; the five labelled subtypes are merged into power-angle, voltage and frequency instability, and a class prediction counts as correct when any of its subtypes is predicted; an unstable event may belong to several classes)}
\begin{tabularx}{\textwidth}{>{\RaggedRight}Xll>{\RaggedRight}X}
\toprule
\textbf{Instability class} & \textbf{Recall (\%)} & \textbf{Precision (\%)} & \textbf{Share of unstable events (\%)} \\
\midrule
Power-angle instability & 88.2 [83.4--91.2] & 92.9 [89.0--95.2] & 57.2 \\
Voltage instability (undervoltage or overvoltage) & 99.2 [96.7--99.9] & 93.1 [92.3--94.1] & 92.3 \\
Frequency instability (low or high frequency) & 98.3 [96.8--98.9] & 95.9 [94.0--97.5] & 87.9 \\
\bottomrule
\end{tabularx}
\end{table}

\subsection*{3.4. Sensitivity of transient-stability assessment to encoder adaptation}

Keeping the grid encoder frozen allows the same representation mechanism to be reused across different grid structures and computational tasks, but it may also restrict task-specific optimization. We therefore examined whether allowing the encoder to adapt to TSA produced a measurable improvement in classification performance on IEEE118. Three configurations were compared: a frozen encoder, for which no encoder parameters were updated during TSA training; partial adaptation, for which only the final two encoder layers were trainable; and full adaptation, for which all encoder layers were trainable. All three configurations used the same TSA task formulation. TSA accuracy was defined as the proportion of fault events for which the predicted stability outcome agreed with the corresponding time-domain simulation result.

We used three complementary comparison protocols because encoder adaptation can affect both the learning trajectory and the stopping point. The first was an early fixed-budget comparison, in which all three configurations were evaluated after exactly 4,000 optimization steps; this provides a comparison at the same early training budget, although all three configurations were still undertrained at this stage. The second was a matched-checkpoint comparison between the frozen and fully adapted configurations: five paired TSA training runs were evaluated at the same three checkpoints 6,200, 7,200 and 8,200 optimization steps, and the full-minus-frozen accuracy difference is reported at each checkpoint separately. The third comparison used the independently selected stopping point of each configuration; it reflects the selected endpoint of each training run, but is less controlled because the frozen and fully adapted models may have undergone different numbers of optimization steps.

A positive value indicates higher TSA accuracy with full encoder adaptation. Differences are reported in percentage points (pp) and represent absolute changes in accuracy rather than relative percentage improvements.

\begin{table}[htbp]
\centering
\footnotesize
\caption*{Table 14 | Sensitivity of transient-stability assessment to encoder adaptation (IEEE118 accuracy). The 4,000-step row is the mean of three seeds at a fixed early budget; the checkpoint rows are means of five paired training runs evaluated at the same optimization step; the last row compares the independently selected stopping point of each run. These sensitivity runs use a held-out split and training schedule different from the five realizations of Table 12}
\begin{tabularx}{\textwidth}{>{\RaggedRight}Xl>{\RaggedRight}Xll}
\toprule
\textbf{Comparison point} & \textbf{Frozen encoder} & \textbf{Final two layers adapted} & \textbf{Fully adapted} & \textbf{Full - frozen (pp)} \\
\midrule
Fixed budget at 4,000 steps (three seeds, mean) & 0.8488 & 0.8465 & 0.8459 & -0.29 \\
Matched checkpoint at 6,200 steps (five pairs, mean) & 0.9060 & --- & 0.9116 & +0.57 \\
Matched checkpoint at 7,200 steps & 0.9179 & --- & 0.9234 & +0.55 \\
Matched checkpoint at 8,200 steps & 0.9208 & --- & 0.9285 & +0.77 \\
Independently selected stopping points (mean 9,000 steps frozen, 12,040 full) & 0.9236 & --- & 0.9529 & +2.93 \\
\bottomrule
\end{tabularx}
\end{table}

At the fixed 4,000-step training budget, TSA accuracy was 0.8488 with the encoder frozen, 0.8465 when only the final two encoder layers were adapted and 0.8459 with full encoder adaptation (Table 14). The differences were small, but all three configurations remained undertrained at this stage. The 4,000-step experiment therefore provides an early training-trajectory comparison rather than evidence that one adaptation strategy was superior.

The matched-checkpoint comparison shows the same picture at every anchor: full adaptation led freezing by 0.57 pp at 6,200 steps, 0.55 pp at 7,200 steps and 0.77 pp at 8,200 steps on average, and the difference was positive in three of the five paired runs at each checkpoint. At a matched training length, the advantage of full adaptation therefore stayed below one percentage point and was not reproduced in all runs. When the two configurations were instead compared at their independently selected stopping points, the full-adaptation advantage grew to 2.93 pp and was positive in all five runs; the fully adapted runs, however, also trained longer before stopping: 12,040 optimization steps on average against 9,000 for the frozen runs (individual stopping points 6,600--15,600), so this larger difference cannot be separated from the effect of training duration and is treated as secondary evidence.

Under matched training conditions, the frozen encoder therefore reached TSA accuracy within one percentage point of full task-specific adaptation on average, and the small advantage of adaptation was not reproduced in all five training runs. This conclusion is specific to transient-stability assessment and does not imply that freezing is generally preferable to task-specific adaptation. Rather, for the tested TSA setting, maintaining the shared encoder unchanged resulted in performance close to that obtained with full task-specific adaptation.

\section*{Discussion}

Our results support the feasibility of a unified representation of structurally different power grids through a common latent description learned without downstream task supervision. The purpose of the encoder is to provide a faithful and structurally consistent description of the grid itself. We therefore evaluated the encoder primarily by information recoverability, that is, how well the physical and operating information encoded in the latent space could be reconstructed as network structure and scale changed. Once encoded, each grid is represented by a variable number of fixed-dimensional node, branch and global vectors. Grids of different sizes thus retain their own system-specific content while sharing the same representation format, latent dimensionality and token semantics. In this sense, each grid has its own structured latent description, while all grids follow the same representation scheme. Downstream reasoners are then trained to interpret this fixed representation and extract the information required for their respective computational objectives, rather than redefining the grid representation for each task. This separates the representation of the physical grid from task-specific computation.

The effectiveness of our architecture arises not from freezing alone, but from a division of labour among representation learning, task-specific reasoning and explicit physical information.\textbf{ }Multi-field masked pretraining organizes grid topology, equipment attributes, operating states and element identity into aligned node, edge and global tokens. During downstream learning, the encoder weights, normalization statistics and token semantics remain fixed, while independently trained query reasoners select and recombine the information required by each task. Task-specific quantities including known power-flow variables, violation magnitudes, control constraints and fault conditions are supplied outside the encoder, and physics-based hints, correction procedures or simulation checks provide the accuracy and feasibility required by individual applications. This arrangement allows the encoder to serve as a stable, element-addressable coordinate system rather than requiring it to solve every downstream problem by itself. The distinction from previous power-system foundation models is therefore not the use of a frozen backbone per se: frozen-encoder transfer has already been explored for power-system time-series tasks and within specialized task families, while unified neural solvers have coupled shared latent states to task supervision or iterative physics feedback\textsuperscript{[3,9,10,22]}. Instead, our contribution is the combined use of a permanently fixed grid representation with fixed normalization and node, directed-edge and global-token semantics, and external task reasoners across grid structures and heterogeneous objectives spanning steady-state calculation and control, operating-condition generation and transient-stability assessment. Element-level addressability avoids the information loss associated with compressing a grid into a single pooled vector; keeping the shared coordinate system fixed limits its distortion by task-specific gradients; and supplying task conditions externally avoids forcing one backbone to encode every future objective. The observed performance should therefore be attributed to the cooperation of the frozen representation, selective task readout and physics-aware modules, rather than to the encoder alone.

The different recoverability of voltage magnitude and phase angle is consistent with their different physical roles in AC power systems. Classical power-flow analysis establishes a strong \textit{P}--\textit{$\theta$} coupling and \textit{Q} --\textit{V} coupling, with comparatively weak cross-coupling between the two subsystems\textsuperscript{[}\textsuperscript{48}\textsuperscript{]}. Voltage magnitude is strongly related to local and neighbouring reactive-power balance, voltage regulation, and the surrounding load and generation states\textsuperscript{[19]}. Much of this information is directly available to the encoder, providing multiple complementary cues for reconstructing a masked voltage magnitude. Phase angle, in contrast, is more strongly determined by system-wide active-power transfer and the electrical coupling imposed by network parameters\textsuperscript{[21]}. Recovering the angle at a given bus therefore requires information associated with power-transfer paths that can extend well beyond its immediate neighbourhood. This difference is reflected in the reconstruction results. The stronger degradation of phase-angle recovery with increasing grid scale is also consistent with its greater dependence on network-wide information.

The present results also indicate several directions for further development. First, encoder pretraining was deliberately restricted to small- and medium-scale power systems, while substantially larger networks were reserved for transfer evaluation. This setting introduced a clear scale shift and allowed the generalization capability of the representation to be examined beyond the training range. Future work should extend encoder pretraining to a larger and more diverse set of network structures and system scales, including large-scale transmission systems, and evaluate how the unified representation improves as the pretraining coverage is expanded. Second, the physical information represented by the encoder can be further enriched. The relatively weak recoverability of phase angle motivates improved angle representations, such as relative bus angles or branch angle differences that are less dependent on the choice of angular reference. The grid description should also be extended to hybrid AC/DC networks and include a more complete set of dynamic model parameters for generators, converters, controllers and protection systems. Third, the computational architecture could be extended from unified representation to unified end-to-end computation. In the present framework, the grid is encoded once, but intermediate results are still returned to explicit power-system variables before being passed to subsequent computations. A more general architecture could keep these intermediate states in the shared latent space and propagate them directly between computational modules. The physical grid would then need to be encoded only once at the beginning of the computational chain. Subsequent steady-state analysis, dynamic assessment, control and generation tasks could operate on and update the same latent representation, with explicit physical quantities decoded only when required as final outputs. Such a formulation would move beyond reuse of a common grid representation toward a unified computational model, in which multiple power-system calculations are composed end to end within a common latent space. A first step in this direction is already present in our framework: the operating-condition generator does not merely read the latent representation as a fixed input; it synthesizes new operating states directly in the shared latent space, generating latent node states conditioned on the requested operating condition and decoding them into explicit grid quantities only for delivery and physical verification. Extending this single-module latent-space generation to computations chained across multiple reasoners remains future work

\section*{Encoder availability}

The pretrained weights of the unified grid encoder are available on GitHub at \url{https://github.com/goldenlcq-debug/gridfm-encoder-v1}.

\section*{References}
\begingroup
\renewcommand{\section}[2]{}

\endgroup


\begin{thebibliography}{48}

\bibitem{ref1} Li, Y. et al. Artificial intelligence-based methods for renewable power system operation. Nat. Rev. Electr. Eng. 1, 163--179 (2024).

\bibitem{ref2} Hamann, H. F. et al. Foundation models for the electric power grid. Joule 8, 3245--3258 (2024).

\bibitem{ref3} Puech, A. et al. GENCO---A unified neural solver embedded in a development framework for steady-state grid analysis. Preprint at \url{https://doi.org/10.48550/arXiv.2608.09921} (2026).

\bibitem{ref4} Varbella, A., Amara, K., Gjorgiev, B., El-Assady, M. \& Sansavini, G. PowerGraph: a power grid benchmark dataset for graph neural networks. Adv. Neural Inf. Process. Syst. 37 (2024).

\bibitem{ref5} Gillioz, M., Dubuis, G. \& Jacquod, P. A large synthetic dataset for machine learning applications in power transmission grids. Sci. Data 12, 168 (2025).

\bibitem{ref6} Puech, A., Weiss, J., Brunschwiler, T. \& Hamann, H. F. Optimal Power Grid Operations with Foundation Models. Preprint at \url{https://arxiv.org/abs/2409.02148} (2024).

\bibitem{ref7} Li, Y. et al. LUMINA: foundation models for topology transferable ACOPF. Preprint at \url{https://doi.org/10.48550/arXiv.2603.04300} (2026).

\bibitem{ref8} Zhu, Y. et al. GNNs' generalization improvement for large-scale power system analysis based on physics-informed self-supervised pre-training. IEEE Trans. Power Syst. 40, 4145--4157 (2025).

\bibitem{ref9} Papaioannou, C. et al. MxGPS: multiplex graph transformers for a power grid foundation model. Preprint at \url{https://doi.org/10.48550/arXiv.2607.13763} (2026).

\bibitem{ref10} Lupo Pasini, M., Li, Y., Kim, K. \& Kuruganti, T. Scalable heterogeneous graph foundation models for data-driven optimal power flow in smart grids. Preprint at \url{https://doi.org/10.48550/arXiv.2605.23194} (2026).

\bibitem{ref11} Arowolo, O. \& Cremer, J. L. Towards generalization of graph neural networks for AC optimal power flow. Preprint at \url{https://doi.org/10.48550/arXiv.2510.06860} (2025).

\bibitem{ref12} Wu, T., Scaglione, A., Miguel, S. \& Arnold, D. Universal graph learning for power system reconfigurations: transfer across topology variations. Preprint at \url{https://doi.org/10.48550/arXiv.2509.08672} (2025).

\bibitem{ref13} Lovett, S. et al. OPFData: large-scale datasets for AC optimal power flow with topological perturbations. Preprint at \url{https://doi.org/10.48550/arXiv.2406.07234} (2024).

\bibitem{ref14} Piloto, L. et al. CANOS: a fast and scalable neural AC-OPF solver robust to N-1 perturbations. Preprint at \url{https://doi.org/10.48550/arXiv.2403.17660} (2024).

\bibitem{ref15} Lin, N., Orfanoudakis, S., Ordonez Cardenas, N., Giraldo, J. S. \& Vergara, P. P. PowerFlowNet: power flow approximation using message passing graph neural networks. Int. J. Electr. Power Energy Syst. 160, 110112 (2024).

\bibitem{ref16} Li, Y. et al. Scale-adaptive power flow analysis with local topology slicing and multi-task graph learning. Preprint at \url{https://doi.org/10.48550/arXiv.2601.01387} (2026).

\bibitem{ref17} Nakiganda, A. M. \& Chatzivasileiadis, S. Graph neural networks for fast contingency analysis of power systems. Preprint at \url{https://doi.org/10.48550/arXiv.2310.04213} (2023).

\bibitem{ref18} de Jong, M., Viebahn, J. \& Shapovalova, Y. Graph neural networks for transmission grid topology control: busbar information asymmetry and heterogeneous representations. Preprint at \url{https://doi.org/10.48550/arXiv.2501.07186} (2025).

\bibitem{ref19} Simpson-Porco, J. W., D\"orfler, F. \& Bullo, F. Voltage collapse in complex power grids. Nat. Commun. 7, 10790 (2016).

\bibitem{ref20} Sch\"afer, B. et al. Understanding Braess' paradox in power grids. Nat. Commun. 13, 5396 (2022).

\bibitem{ref21} D\"orfler, F., Chertkov, M. \& Bullo, F. Synchronization in complex oscillator networks and smart grids. Proc. Natl Acad. Sci. USA 110, 2005--2010 (2013).

\bibitem{ref22} Tu, S. et al. PowerPM: foundation model for power systems. Adv. Neural Inf. Process. Syst. 37, 115233--115260 (2024).

\bibitem{ref23} Stiasny, J. \& Cremer, J. Residual power flow for neural solvers. Preprint at \url{https://doi.org/10.48550/arXiv.2601.09533} (2026).

\bibitem{ref24} Mohammadian, M., Van Boven, A. \& Baker, K. Restoring feasibility in power grid optimization: a counterfactual ML approach. Preprint at \url{https://doi.org/10.48550/arXiv.2504.06369} (2025).

\bibitem{ref25} Bohigas-Daranas, F., Latif-Martinez, H., Prieto-Araujo, E., Barlet-Ros, P. \& Gomis-Bellmunt, O. Power flow feasibility assessment using variational graph autoencoders. Preprint at \url{https://doi.org/10.48550/arXiv.2607.09122} (2026).

\bibitem{ref26} Xiao, C., He, X., Li, H., Tong, H. \& Weng, Y. Operationally feasible synthetic power-grid scenarios via learning the AC-operable joint distribution. Preprint at \url{https://doi.org/10.48550/arXiv.2608.03878} (2026).

\bibitem{ref27} He, X. et al. PowerGrow: feasible co-growth of structures and dynamics for power grid synthesis. Preprint at \url{https://doi.org/10.48550/arXiv.2509.12212} (2025).

\bibitem{ref28} Hoseinpour, M. \& Dvorkin, V. Constrained diffusion models for synthesizing representative power flow datasets. Preprint at \url{https://doi.org/10.48550/arXiv.2506.11281} (2025).

\bibitem{ref29} Wang, J., Upadhyay, D., Zaman, M. \& Srikantha, P. Synthetic power flow data generation using physics-informed denoising diffusion probabilistic models. Preprint at \url{https://doi.org/10.48550/arXiv.2504.17210} (2025).

\bibitem{ref30} Chen, Y., Wang, Y., Kirschen, D. \& Zhang, B. Model-free renewable scenario generation using generative adversarial networks. IEEE Trans. Power Syst. 33, 3265--3275 (2018).

\bibitem{ref31} Shen, C., Zuo, K. \& Sun, M. Universal transient stability analysis: a pre-trained generative Transformer-enabled power system dynamics prediction framework. Preprint at \url{https://doi.org/10.48550/arXiv.2512.20970} (2025).

\bibitem{ref32} Li, H., Mai, L., Xiao, C., Blasch, E. \& Weng, Y. Predicting power-system dynamic trajectories with foundation models. Preprint at \url{https://doi.org/10.48550/arXiv.2604.14991} (2026).

\bibitem{ref33} Arowolo, O., Yang, M. \& Cremer, J. Revisiting data-driven dynamic security assessment with a tabular foundation model. Preprint at \url{https://doi.org/10.48550/arXiv.2607.16031} (2026).

\bibitem{ref34} Yang, S.-G., Kim, B. J., Son, S.-W. \& Kim, H. Power-grid stability predictions using transferable machine learning. Chaos 31, 123127 (2021).

\bibitem{ref35} Nauck, C. et al. Predicting basin stability of power grids using graph neural networks. New J. Phys. 24, 043041 (2022).

\bibitem{ref36} Nauck, C., Lindner, M., Sch\"urholt, K. \& Hellmann, F. Toward dynamic stability assessment of power grid topologies using graph neural networks. Chaos 33, 103103 (2023).

\bibitem{ref37} Fan, S. et al. Review on data-driven power system transient stability assessment technology. Proc. CSEE 44, 3408--3428 (2024).

\bibitem{ref38} Guo, Q. et al. Architecture and key technologies of hybrid-intelligence-based decision-making of operation modes for new type power systems. Electric Power 56, 1--13 (2023).

\bibitem{ref39} Thams, F., Venzke, A., Eriksson, R. \& Chatzivasileiadis, S. Efficient database generation for data-driven security assessment of power systems. IEEE Trans. Power Syst. 35, 30--41 (2020).

\bibitem{ref40} Giraud, B., Charles, L., Nakiganda, A. M., Vorwerk, J. \& Chatzivasileiadis, S. A dataset generation toolbox for dynamic security assessment: on the role of the security boundary. Sustain. Energy Grids Netw. 43, 101833 (2025).

\bibitem{ref41} Yang, W. et al. GridSFM: a foundation model for AC optimal power flow. Microsoft Research Technical White Paper (2026).

\bibitem{ref42} Bommasani, R. et al. On the opportunities and risks of foundation models. Preprint at \url{https://doi.org/10.48550/arXiv.2108.07258} (2021).

\bibitem{ref43} Hou, Z. et al. GraphMAE: self-supervised masked graph autoencoders. In Proc. 28th ACM SIGKDD Conf. Knowledge Discovery and Data Mining 594--604 (ACM, 2022).

\bibitem{ref44} You, Y. et al. Graph contrastive learning with augmentations. Adv. Neural Inf. Process. Syst. 33, 5812--5823 (2020).

\bibitem{ref45} Ramp\'a\v{s}ek, L. et al. Recipe for a general, powerful, scalable graph Transformer. Adv. Neural Inf. Process. Syst. 35, 14501--14515 (2022).

\bibitem{ref46} Lake, B. M. \& Baroni, M. Generalization without systematicity: on the compositional skills of sequence-to-sequence recurrent networks. Proc. Mach. Learn. Res. 80, 2873--2882 (2018).

\bibitem{ref47} Zhou, K., Liu, Z., Qiao, Y., Xiang, T. \& Loy, C. C. Domain generalization: a survey. IEEE Trans. Pattern Anal. Mach. Intell. 45, 4396--4415 (2023).

\bibitem{ref48} Stott, B. \& Alsac, O. Fast decoupled load flow. IEEE Trans. Power Appar. Syst. PAS-93, 859--869 (1974).

\end{thebibliography}
\end{document}